\documentclass[useAMS,usenatbib]{mn2e}
\usepackage{graphicx}
\usepackage{natbib}
\usepackage[para]{threeparttable}
\usepackage{xcolor}
\usepackage{hyperref}
\usepackage{mhchem}

\usepackage{todonotes}
\def\lesssim{\mathrel{\hbox{\rlap{\hbox{\lower3pt\hbox{$\sim$}}}\hbox{\raise2pt\hbox{$<$}}}}}
\def\gtrsim{\mathrel{\hbox{\rlap{\hbox{\lower3pt\hbox{$\sim$}}}\hbox{\raise2pt\hbox{$>$}}}}}
\newcommand\ion[2]{#1$\;${\scshape{#2}}} 
\title[Direct Analysis of SN~2012dn Optical Spectra]{Comparative Analysis of SN~2012dn Optical Spectra: \\ Days $-$14 to $+$114}
\author[J. T. Parrent]
{\parbox{\textwidth}{J. T. Parrent$^{1,2,3}$\thanks{E-mail: jparrent@cfa.harvard.edu (JTP)}, D.~A.~Howell$^{2,4}$, R.~A.~Fesen$^{3}$, S.~Parker$^{5}$, F.~B.~Bianco$^{6}$, B.~Dilday$^{7}$, D.~Sand$^{8}$, S.~Valenti$^{2,3}$, J.~Vink\'o$^{9,10}$, P.~Berlind$^{11}$, P.~Challis$^{1}$, D.~Milisavljevic$^{1}$, N.~Sanders$^{1}$, G.~H.~Marion$^{10}$, J.~C.~Wheeler$^{10}$, P.~Brown$^{12}$, M.~L.~Calkins$^{6}$, B.~Friesen$^{13}$, R.~Kirshner$^{1}$, T.~Pritchard$^{14}$, R.~Quimby$^{15,16}$, and P.~Roming$^{14,17}$\\
}\vspace{0.4cm}\\
\parbox{\textwidth}
{
$^{1}$Harvard-Smithsonian Center for Astrophysics, 60 Garden St., Cambridge, MA 02138, USA\\
$^{2}$Las Cumbres Observatory Global Telescope Network, Goleta, CA 93117, USA\\
$^{3}$6127 Wilder Lab, Department of Physics \& Astronomy, Dartmouth College, Hanover, NH 03755, USA\\
$^{4}$Department of Physics, U.C. Santa Barbara, Santa Barbara, CA 93117, USA\\
$^{5}$Backyard Observatory Supernova Search, Parkdale Observatory, Canterbury 7495, New Zealand\\
$^{6}$Center for Cosmology and Particle Physics, New York University, 4 Washington Place, New York, NY 10003, USA\\
$^{7}$Gravity Jack, 23505 E Appleway Ave \#200, Liberty Lake, WA 99019, USA\\
$^{8}$Physics Department, Texas Tech University, Lubbock, TX 79409, USA\\
$^{9}$Department of Optics and Quantum Electronics, University of Szeged, D\'{o}m t\'{e}r 9, 6720 Szeged, Hungary\\
$^{10}$Department of Astronomy, University of Texas, Austin, TX 78712, USA\\
$^{11}$F. L. Whipple Observatory, 670 Mt. Hopkins Road, P.O. Box 97, Amado, AZ 85645, USA\\
$^{12}$Department of Physics and Astronomy, George P. and Cynthia Woods Mitchell Institute for Fundamental Physics \& Astronomy, Texas A. \& M. University, 4242 TAMU, College Station, TX 77843, USA\\
$^{13}$Homer L. Dodge Department of Physics and Astronomy, University of Oklahoma, 440 W Brooks, Norman, OK 73019, USA\\
$^{14}$Department of Astronomy \& Astrophysics, Penn State University, 525 Davey Lab, University Park, PA 16802, USA\\
$^{15}$Department of Astronomy, San Diego State University, 5500 Campanile
Drive, San Diego, CA 92182-1221\\
$^{16}$Kavli IPMU (WPI), UTIAS, The University of Tokyo, Kashiwa, Chiba 277-8583, Japan\\
$^{17}$Southwest Research Institute, Department of Space Science, 6220 Culebra Road, San Antonio, TX 78238, USA
}}

\begin{document}

\pagerange{\pageref{firstpage}--\pageref{lastpage}} \pubyear{2015}

\maketitle

\label{firstpage}

\begin{abstract} 

SN~2012dn is a super-Chandrasekhar mass candidate in a purportedly normal spiral (SAcd) galaxy, and poses a challenge for theories of type Ia supernova diversity. Here we utilize the fast and highly parameterized spectrum synthesis tool, \texttt{SYNAPPS}, to estimate relative expansion velocities of species inferred from optical spectra obtained with six facilities. 
As with previous studies of normal SN~Ia, we find that both unburned carbon and intermediate mass elements are spatially coincident within the ejecta near and below 14,000~km~s$^{-1}$. 
Although the upper limit on SN~2012dn's peak luminosity is comparable to some of the most luminous normal SN~Ia, we find a progenitor mass exceeding $\sim$~1.6~M$_{\odot}$ is not strongly favored by leading merger models since these models do not accurately predict spectroscopic observations of SN~2012dn and more normal events. 
 In addition, a comparison of light curves and host-galaxy masses for a sample of literature and Palomar Transient Factory SN~Ia reveals a diverse distribution of SN~Ia subtypes where carbon-rich material remains unburned in some instances. Such events include SN~1991T, 1997br, and 1999aa where trace signatures of \ion{C}{III} at optical wavelengths are presumably detected.

\end{abstract}


\clearpage

\section{Introduction}

The progenitor systems of type~Ia supernovae (SN~Ia) have not been observed, although it is widely believed that some SN~Ia are the product of a thermonuclear runaway in a white dwarf star (WD) with a mass near the Chandrasekhar limit, M$_{Ch}$~$\approx$~1.38~M$_{\odot}$ (\citealt{Nugent11} and references therein). There is also observational evidence to suggest that the progenitor mass for a fraction of over-luminous events may exceed M$_{Ch}$ \citep{Howell06,Jeffery06,Yamanaka09a,Scalzo10}, while the progenitor mass for some fraction of normal and sub-luminous SN~Ia may be notably less than M$_{Ch}$ \citep{Woosley11,Shen14,Scalzo14subchandra, Dan15}. 

Studies of SN~Ia therefore face an interesting challenge as these exploded WD systems may be the merger of two sub-M$_{Ch}$ WDs \citep{Kromer13SN2010lp}, some involving low-mass helium stars \citep{Shen14,Dan15}; coalesced mergers or prompt ``peri-mergers'' of two near-M$_{Ch}$ WDs \citep{Moll14,Raskin14}; massive WDs in single-degenerate systems \citep{Hachisu12a,Chen14}; and single-degenerate systems with sub-M$_{Ch}$ ejecta masses \citep{Chiosi15}. Spectroscopic modeling has also yet to distinguish appearances between single and double-degenerate scenarios \citep{Liu97,KasenPlewa07,Roepke12,Mazzali15,Dessart14models}. The uniqueness of progenitor systems that produce super-M$_{Ch}$ mass candidates (SCC) and more common SN~Ia thus remains an open issue.

Early-phase optical spectra of the SCC class are thus far characterized by relatively weak and narrow spectral features, moderate to low mean expansion velocities (8~$-$~16~x~10$^{3}$~km~s$^{-1}$), lower than normal line-velocity gradients of blended absorption minima, and higher than normal integrated flux near maximum light \citep{Scalzo12,Brown14}. This in combination with relatively large and long-lasting signatures of \ion{C}{II}~$\lambda\lambda$~6580,~7234 suggest the presence of higher than normal amounts of unburned carbon \citep{Howell06,Scalzo10,Taubenberger13}. Although there are few well-observed SCC SN Ia, e.g., the over-luminous SN~2009dc \citep{Yamanaka09a,Tanaka10,Silverman11}, the most recent, SN~2012dn, reached a peak absolute magnitude that was not exceptionally luminous \citep{Brown14,Chakradhari14}, and the same can be said for the spectroscopically similar SN~2006gz \citep{Maeda09}. 

For some SCC SN~Ia, signatures of \ion{C}{II} may not be conspicuous above low signal-to-noise ratios, or the signature may be fast-evolving similar to that seen for the normal, SN~1999aa-like hybrid, SN~2013dy (see also SN~1999ac; \citealt{Garavini05}). Signatures of carbon might also be detected as \ion{C}{III}~$\lambda$4649 in so-called ``Shallow Silicon'' SN~Ia that display spectral signatures of overall higher ionization species, e.g., SN~1991T and 1997br \citep{Hatano02,Garavini04,Parrent11,Zheng13,Scalzo14}.  Seemingly smaller amounts of unburned material of the C$+$O progenitor, or burned material of the He$+$C$+$O progenitor, are frequently detected for so-called ``Core Normal'' SN~Ia \citep{Jeffery92,Fisher97,Branch05,Thomas07}, while signatures of \ion{C}{I}~$\lambda$10691 are detected for several events, and possibly all SN~Ia \citep{Hoflich02,Marion15,Hsiao15}. 

Signatures of carbon in the outermost ejecta may be a general characteristic of Pulsational-Delayed Detonation(-like) explosion mechanisms (PDD), as proposed for SN~2011fe and 2013dy-like events \citep{Khokhlov91PRD,Bravo09,Dessart14models}, in addition to peculiar SN~2002cx-like events \citep{Baron14}. Carbon-rich regions may also be associated with the remains of a degenerate secondary star that is distributed amorphously at the time of the explosion, where signatures of \ion{C}{II} may not be physically associated with \ion{C}{I} deeper within the ejecta of the primary WD (cf. \citealt{Taubenberger11,Taubenberger13,Moll14,Raskin14}). Adding to this complexity is whether normal SN~Ia originate from a sub-Chandrasekhar mass primary WD \citep{Woosley94,Liu98,Kerkwijk10,Pakmor12,Raskin13,Scalzo14subchandra,Chiosi15}. 

From comparisons of synthesized photometry and spectra, \citet{Pakmor12} found moderate agreement between the normal SN~Ia~2003du and their violent merger model totaling 2.0~M$_{\odot}$ (see also \citealt{Moll14}). \citet{Raskin14} later presented four post-merger models, showing that the angle-dependent, maximum-light spectra of their most normal SN~Ia-like 0.9~$+$~0.8~M$_{\odot}$ model are not also well-matched to most SN~Ia belonging to the SCC class (see also \citealt{vanRossum15}). More recently, \citet{Dan15} reported on their post-merger remnant calculations from an initial sample of nine WD configurations. Out of the three models in fair agreement with spectroscopic abundance estimates of ``normal'' SN~Ia, the total masses for those models are 1.6~M$_{\odot}$ and 2.1~M$_{\odot}$, where the former involves a 0.45~M$_{\odot}$ helium WD. A common theme for these WD merger models is that the overlap between the bulk of the carbon and intermediate-mass ejecta appears to be dependent on the total progenitor mass and viewing angle. 

In terms of observed line velocities, carbon-rich regions of absorbing material for both normal and SCC SN~Ia are estimated to be spatially coincident with freshly synthesized intermediate-mass elements, e.g., Mg, Si, S, and Ca \citep{Branch05,Hicken07}. However, the measurable degree of overlap between potentially unburned progenitor material and regions of Mg, Si, and Ca remains uncertain on account of excessive line-blending \citep{vanRossum12,Foley13profiles,REVIEW}. There is also evidence to suggest the presence of multiple high velocity shells of \ion{Ca}{II} in the outermost ejecta \citep{Thomas04}, which is more than the single high velocity component typically assumed for surveys of SN~Ia spectra (e.g., \citealt{Silverman13SN2013bh,Childress14,Maguire14,Pan15}).

In this article we supplement previously collected data and analyse our time-series optical observations of SN~2012dn during its photospheric phases, starting approximately 15 days prior to maximum light, with late-time coverage at $\sim$~114 days after peak brightness. In \S2 we present our collected {\it gri} photometry and optical spectra. In \S3 we assess the peak absolute magnitude of SN~2012dn and compare to light curves of other SN~Ia prototypes. In \S4 we examine dissimilarities between the spectra of SN~2012dn and literature SN~Ia. In \S5 we outline our approach in utilizing the fast and highly parameterized LTE line-identification tool, \texttt{SYNAPPS} \citep{ThomasSYNAPPS}. In \S6 we analyse the spectra of SN~2012dn with \texttt{SYNAPPS} and discuss our application of a simplified line-identification schema. In \S7 we overview SN~2012dn within the greater context of SN~Ia diversity. In \S8 we summarize and highlight our findings. 

\begin{table*}
 \centering
 \begin{minipage}{140mm}
 \centering
  \caption{FTS Photometry for SN~2012dn}
  \begin{tabular}{lllccc}
  \hline
 UT & MJD & Phase & SDSS-g & SDSS-r & SDSS-i \\
Date & 56000$+$ & (days) & (mag) & (mag) & (mag) \\
\hline
  2012/07/16	& 124.8	& $-$8.0        &  ...	 & 14.543	  (004)	        & 14.863	  (004) 	\\
  2012/07/18	& 126.5	& $-$6.3 & 14.486	  (009)	        & 14.381	  (004)      &14.743	  (004)		\\
  2012/07/19	& 127.5	& $-$5.3 & 14.421  (010)	        & 14.309	  (003)    &14.697	  (004)	\\
  2012/07/20	& 128.6	& $-$4.2 & 14.331	  (010)	        & 14.223	  (003)	       &14.652	  (004)	 	\\
  2012/07/21	& 129.5	& $-$3.3 & 14.322	  (012)	        & 14.199	  (005)       & 14.624	  (003)	 	\\
  2012/07/21	& 129.8	& $-$3.0 & 14.328  (019)	        &14.198	  (009)	          &14.594	  (004)	 	\\
  2012/07/22	& 130.5	& $-$2.3 & 14.297	  (012)	        &14.200	  (007)	     & 14.588	  (004)	 	\\
  2012/07/24	& 132.5	& $-$0.3 & 14.276  (013)	        & 14.085	  (004)	    &14.551	  (006)	 	\\
  2012/07/25	& 133.5	& $+$0.7 & 14.242	  (012)	        &14.131  (008)	       & 14.574  (004)	 	\\
  2012/07/26	& 134.7	& $+$1.9 & 14.249	  (012)	        & 14.119	  (004)	       &14.553	  (003)		\\
  2012/07/28	& 136.5	& $+$3.7 & 14.227	  (053)	        &14.116	  (011)	      & 14.568	  (019)	\\
  2012/07/29	& 137.5	& $+$4.7 & 14.286  (021)	        & 14.123	  (004)	       &14.566	  (005)	 	\\
  2012/07/30	& 138.5	& $+$5.7 & 14.282	  (021)	        &14.124	  (007)	          & 14.569	  (006)		\\
  2012/07/31	& 139.5	  & $+$6.7      &  	... 	 & 14.203	  (009)	        &  	...  	\\
  2012/08/03	& 142.6	& $+$9.8 & 14.528	  (019)	        &  	...        &14.548	  (007)	 	\\
  2012/08/04	& 143.5	& $+$10.8 & 14.492	  (012)	        & 14.220	  (006)       &14.608	  (004)		\\
  2012/08/05	& 144.5	& $+$11.8 & 14.582	  (009)	        & 14.255	  (004)	        & 14.636	  (005)		\\
  2012/08/06	& 145.6	& $+$12.8 & 14.649	  (012)	        & 14.265	  (006)		        & 14.617	  (004)	 	\\
  2012/08/07	& 146.5	& $+$13.7 & 14.690	  (015)	        & 14.312	  (005)    & 14.670	  (006)		\\
  2012/08/09	& 148.5	& $+$15.7 & 14.892	  (010)	        & 14.399	  (004)	       & 14.693	  (004)	 	\\
  2012/08/10	& 149.5	& $+$16.7 & 14.929  (010)        & 14.405	  (003)	      & 14.695	  (004)	 	\\
  2012/08/10	& 149.7	& $+$16.9 & 14.951	  (014)	        & 14.426	  (004)	        & 14.689	  (003)	 	\\
  2012/08/11	& 150.5	& $+$17.7 & 15.065	  (013)	        &14.443	  (004)	        &14.715	  (005)	 	\\
  2012/08/12	& 151.5	& $+$18.7 & 15.148	  (012)	        & 14.475	  (010)	       & 14.738	  (007)	 	\\
  2012/08/12	& 151.7	& $+$18.9 & 15.176	  (012)	        & 14.487	  (004)        &14.708	  (005)	\\
  2012/08/13	& 152.7	& $+$19.9 & 15.333	  (016)	        &14.518	  (007)    &14.713	  (005)\\
  2012/08/14	& 153.5	& $+$20.7 & 15.353	  (013)	        &14.555	  (006)	    & 14.737	  (005)	 \\
  2012/08/15	& 154.5	& $+$21.7 & 15.425	  (020)	        & 14.639	  (015)	     & 14.709	  (005)	 	\\
  2012/08/16	& 155.5	& $+$22.7 & 15.597	  (016)	        &14.601	  (010)	        & 14.751	  (011)	 	\\
  2012/08/20	& 159.5	& $+$26.8 & 15.925	  (034)	        &14.817	  (010)       &14.869	  (013)	\\
  2012/08/21	& 160.5	 & $+$27.8       &  	...   & 14.862	  (016)	        & 14.815	  (098)\\
  2012/08/24	& 163.5	& $+$30.7 & 16.160	  (012)	        &14.988	  (005)	     & 14.981	  (007)	  	\\
  2012/08/30	& 169.5	& $+$36.7 & 16.428	  (028)	        &  15.238	  (009)	   &15.212	  (007)	 	\\
  2012/08/31	& 170.7	& $+$37.9 & 16.458	  (061)	        & 15.285	  (025)	     &15.253	  (020)	\\
  2012/09/01	& 171.5	& $+$38.7 & 16.509	  (021)	        & 15.298	  (006)	       & 15.289	  (006)		\\
  2012/09/02	& 172.5	& $+$38.7 & 16.539  (014)        &15.355	  (007)       &15.395	  (008)	\\
  2012/09/03	& 173.5	& $+$40.7 & 16.590	  (016)	        & 15.404	  (006)	    & 15.386	  (007)	 	\\
  2012/09/04	& 174.5	& $+$41.7 & 16.645	  (016)	        & 15.432	  (007)	      & 15.444	  (005)	  	\\
  2012/09/06	& 176.5	& $+$43.7 & 16.684	  (016)	        & 15.520	  (008)	        &15.491	  (006)	\\
  2012/09/12	& 182.4	& $+$49.6 & 16.836	  (021)	        & 15.778	  (014)	       & 15.691	  (010)	\\
  2012/09/19	& 189.4	& $+$56.6 & 17.036	  (018)	        &16.030	  (015)   & 16.000	  (008)	 	\\
  2012/09/23	& 193.4	& $+$60.6 & 17.134	  (020)	        &16.164	  (019)	    & 16.129	  (012)	 	\\
  2012/09/27	 & 197.4 & $+$64.6 & 17.355	  (051)	        & 16.387	  (032)	      & 16.328	  (013)	  	\\
  2012/09/28	& 198.4	  & $+$65.6      &  	...	 & 16.313	(128)	        & 16.402	  (085)	 	\\
  2012/09/30	& 200.4	& $+$67.6 & 17.384	  (020)	        & 16.487	  (013)      & 16.560	  (010)	\\
  2012/10/04	& 204.4	& $+$71.6 & 17.656	  (033)	        & 16.684	  (019)        & 16.669	  (016)	 	\\
  2012/10/17	& 217.4	& $+$84.6 & 17.848	  (036)	        & 17.204	  (022)	      & 17.264	  (027)	 	\\
  2013/06/22	& 465.7	& $+$332.9 & 20.846	(143)	        & 20.301	  (082)	      &  20.805	(111)	\\
  2013/06/23	& 466.6	& $+$333.8 & 20.953	(195)	        & 20.533	(131)	       &  20.771	(149)	 	\\
  2013/07/07	& 480.6	& $+$347.8 & 19.562	  (050)	        & 19.668	  (055)   & 20.125	  (056)	 	\\
  \hline
\end{tabular}
\end{minipage}
\end{table*}

\section{Observations}

On 2012 July 8 UT an optical transient source with $m_{g}$~$\lesssim$~16 was discovered at $\alpha$(J2000)~=~20$^{h}$23$^{m}$36$^{s}$.26 and $\delta$(J2000) = $-$28$^{o}$16'43''.4 by S.~Parker of the Backyard Observatory Supernova Search team (BOSS; \citealt{Bock12}). We triggered our Gemini-South Target-of-Opportunity programme (GS-2012A-Q-20; PI, D.~A.~Howell) and obtained a first spectrum of SN~2012dn on July 10 UT, which turned out to be approximately 15 days before it reached peak {\it b}-band brightness \citep{Parrent2012dn,Brown14,Chakradhari14}. This spectrum showed weak 6100~\AA\ and 8200~\AA\ features, along with a conspicuous 6300~\AA\ signature, that together suggested SN~2012dn to be a SCC SN~Ia.

\begin{figure*}
\centering
\includegraphics[scale=0.43]{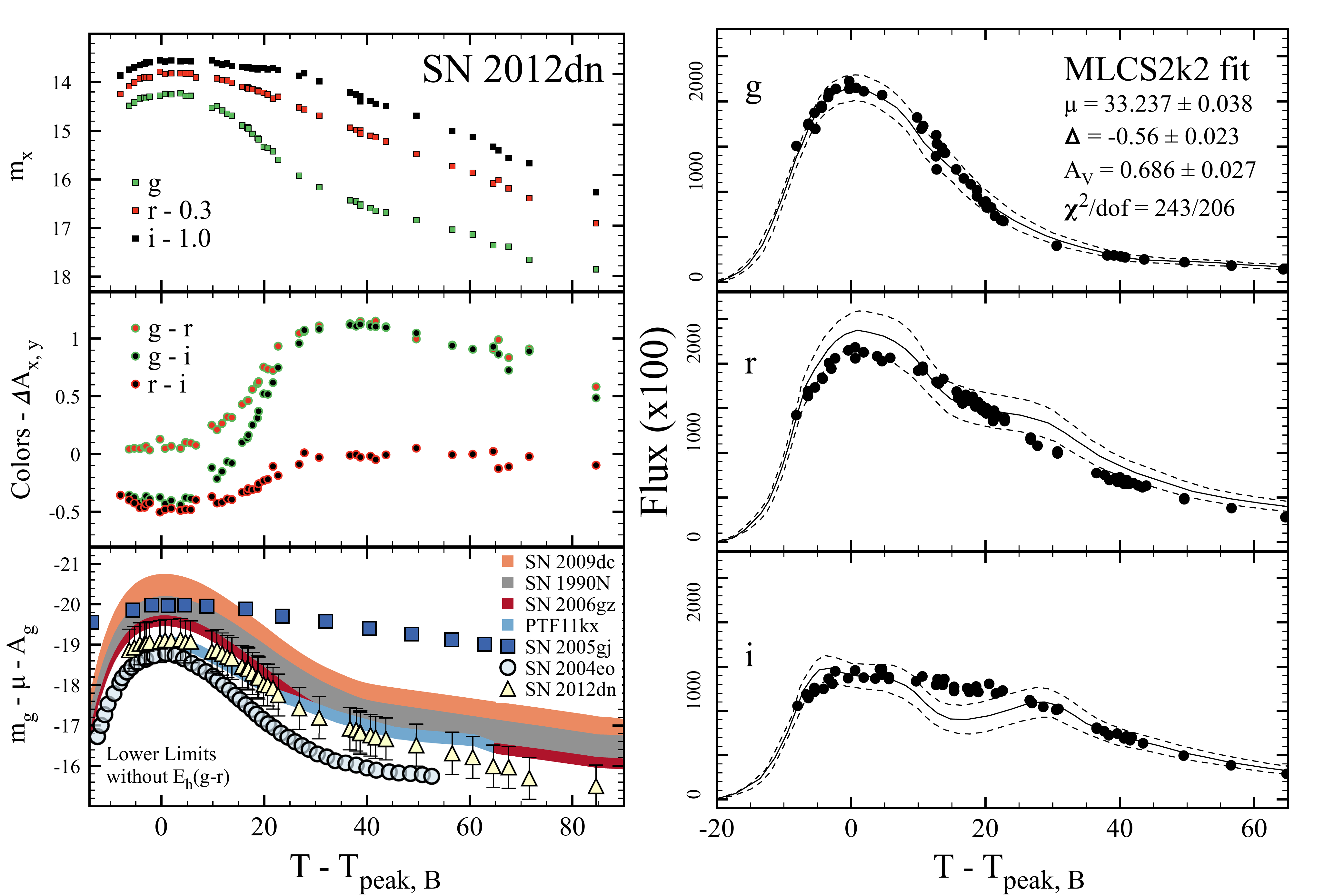}
\caption{{\it Top Left}: FTS {\it gri} observed magnitudes of SN~2012dn during its first 100 days (accounting for $\sim$4 days unobserved prior to day $-$14.6), corrected for MWG extinction, is plotted against \texttt{MLCS2k2} estimated maximum light (MJD 56132.8). {\it Middle Left}: {\it g}$-${\it r}, {\it g}$-${\it i}, and {\it r}$-${\it i} colors of SN~2012dn. {\it Bottom Left}: Absolute {\it g}-band magnitude estimated comparisons of SN~2012dn from \citet{Chakradhari14}, 2006gz, 2009dc, and Core Normal events. Additional photometry was produced with SNLS software to K-correct light curves of SN~1990N \citep{Lira98}, 2006gz \citep{Hicken07}, and 2009dc \citep{Silverman11} from {\it B}, {\it V}, and {\it I}-band to FTS {\it g}, {\it r}, and {\it i}-band filters in the AB photometric system; here we only plot {\it g}-band values. We have supplemented comparisons with light curves for SN~2005gj and PTF11kx (Ia-CSM; \citealt{Aldering06,Prieto07,Dilday12}) and SN~2004eo (an intermediate between normal and SN~1991bg; \citealt{Pastorello07a,Mazzali08}). Error bounds for comparison light curves are with respect to uncertainties in host galaxy extinction and distance modulus. For clarity, we have omitted photometry points at $\sim$ one year post-maximum light. {\it Right} Our \texttt{MLCS2k2} fit to SN~2012dn gri photometry. Filtered photon counts are given on the y-axis, and the x-axis is the time relative to the date of peak brightness. \texttt{MLCS2k2} fitting parameters are given in the upper-right panel.}
\label{Fig:phot}
\end{figure*}

\subsection{Photometry}

Photometric observations using {\it g'r'i'} filters were obtained with the Spectral camera on Faulkes Telescope South (FTS), spanning nearly a year of coverage (see Table~1 and Figure~\ref{Fig:phot}). Our observations began on 2012~July UT 16, about four days after the {\it Swift}/UVOT photometry of \citet{Brown14} and \citet{Chakradhari14}. Images and filtered photon counts were processed and converted to SDSS {\it gri} with an automatic pipeline of standard IRAF procedures (see \citealt{Valenti14,Graham14} for details). 

\subsection{Spectroscopy}

Spectroscopic observations of SN~2012dn were obtained by six facilities: Gemini-North and Gemini-South with GMOS \citep{Hook04,Howell12}, South African Large Telescope (SALT) with RSS \citep{Burgh03,Nordsieck12}, Tillinghast Telescope with the FAST spectrograph \citep{Fabricant98}, Faulkes Telescope South (FTS) with FLOYDS \citep{LCOGT}, and the Multiple Mirror Telescope (MMT) using the Blue Channel instrument \citep{Schmidt89}. Data were reduced with IRAF, including Gemini software add-on packages, and have been corrected for host recessional velocities throughout this manuscript assuming a redshift of z~=~0.010187 \citep{Theureau98}. 

\begin{figure*}
\centering
\includegraphics[scale=0.4, trim= 20mm 65mm 0mm 0mm]{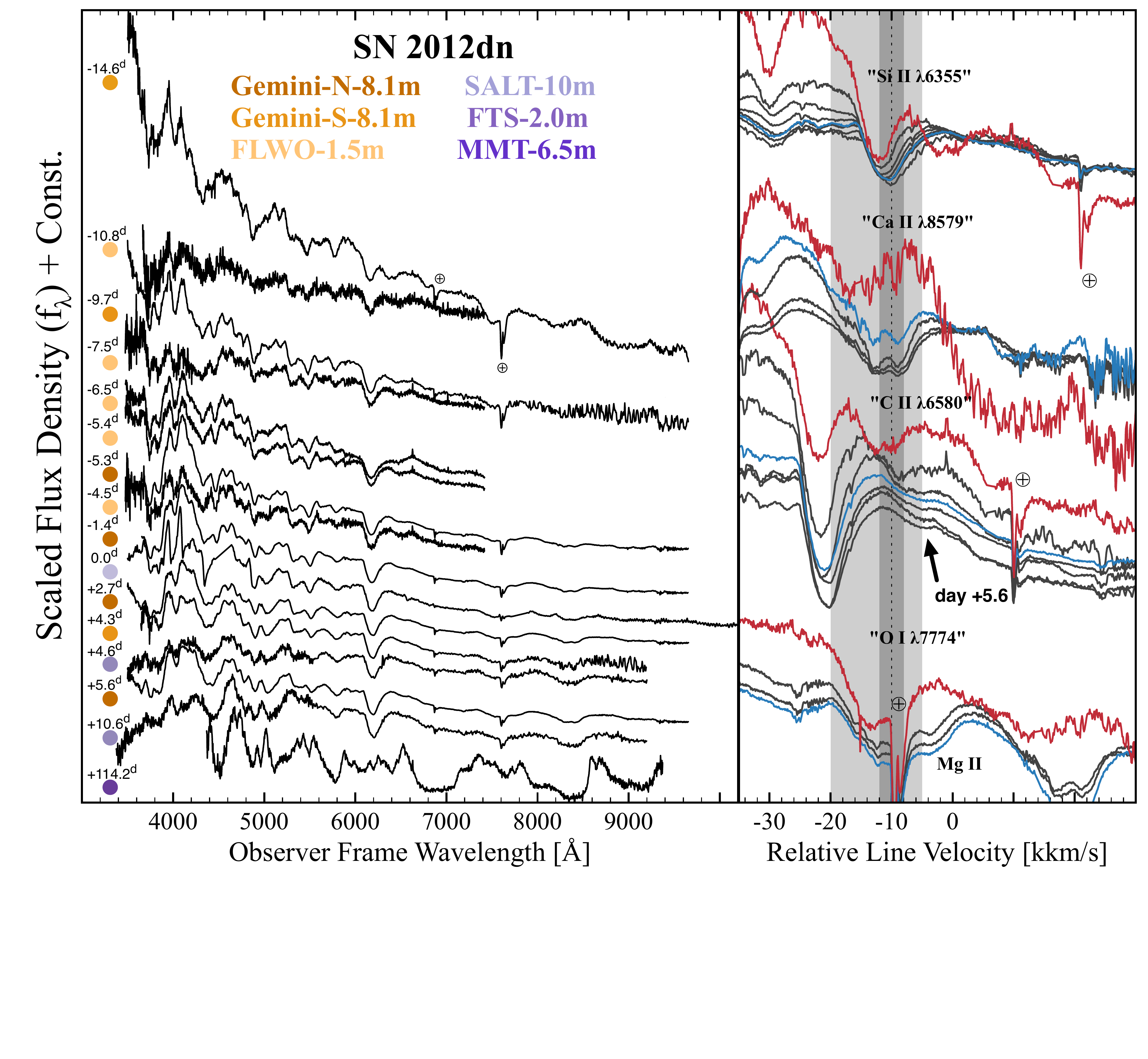}
\caption{{\it Left}: Observer frame spectra for SN~2012dn. Spectra are colored by facilities listed above. Some telluric features are marked by $\oplus$. See Table 2 for details. {\it Right}: Stacked velocity plotted spectra of SN~2012dn pre-maximum light spectra. Spectra are shown with respect to $\mathcal{O}$(v/c) relative line velocities as indicated. See \S5 where our \texttt{SYNAPPS} analysis includes the relativistic correction term for estimating projected Doppler line velocities for rest-frame spectra. Red and blue lines denote the first and maximum light spectrum, respectively. A vertical dark-grey column centered about 10,000~km~s$^{-1}$ is meant to guide the eye for inspecting overlapping line velocities, and the vertical light-grey band spans the approximate width of the 7500~\AA\ feature for reference.}
\label{Fig:specs}
\end{figure*}

For several FAST spectra, galaxy contamination is apparent near 6563~\AA, and telluric features have not been removed in our data set (as indicated in Figure~\ref{Fig:specs}). We note however that neither of these contaminants have significant effect on our subsequent analysis of SN~2012dn spectra in \S6.

\begin{table*}
 \centering
 \begin{minipage}{140mm}
  \centering
  \caption{Time-series Spectra for SN~2012dn}
  \begin{tabular}{cccccc}
  \hline
  UT & MJD & Days Since & Telescope & Exp. Time & Range \\
Date & 56000$+$ & {\it B}-band Maximum & $+$Instrument & (s) & (\AA) \\
\hline
10/07/12 & 118.2 & $-$14.6 & GS$+$GMOS & 450(x2) & 3500$-$9500 \\  
14/07/12 & 122.0 & $-$10.8 & FLWO$+$FAST & 1200 & 3500$-$7400 \\     
15/07/12 & 123.1 & $-$9.7 & GS$+$GMOS & 450(x2) & 3500$-$9500 \\     
17/07/12 & 125.3 & $-$7.5 & FLWO$+$FAST & 870 & 3500$-$7400 \\  
18/07/12 & 126.3 & $-$6.5 & FLWO$+$FAST & 1200 & 3500$-$7400 \\  
19/07/12 & 127.4 & $-$5.4 & FLWO$+$FAST & 1800 & 3500$-$7400 \\  
19/07/12 & 127.5 & $-$5.3 & GN$+$GMOS & 450(x2) & 3500$-$9500 \\ 
20/07/12 & 128.3 & $-$4.5 & FLWO$+$FAST & 700 & 3500$-$7400 \\  
23/07/12 & 131.5 & $-$1.4 & GN$+$GMOS & 450(x2) & 3500$-$9500 \\ 
24/07/12 & 132.8 & 0.0 & SALT & 600 & 3500$-$10000 \\
27/07/12 & 135.5 & $+$2.7 & GN$+$GMOS & 200(x2) & 3500$-$9500 \\
29/07/12 & 137.1 & $+$4.3 & GS$+$GMOS & 120(x2) & 3500$-$9500 \\
29/07/12 & 137.4 & $+$4.6 & FTS$+$FLOYDS & 600 & 3500$-$9100\\
30/07/12 & 138.4 & $+$5.6 & GN$+$GMOS & 240(x2) & 3500$-$9500 \\
04/08/12 &143.4 & $+$10.6 & FTS$+$FLOYDS & 600 & 3500$-$9100 \\
16/11/12 & 247.0 & $+$114.2 & MMT$+$Blue Channel & 900 & 4200$-$9200 \\
\hline
\end{tabular}
\end{minipage}
\end{table*}

Assuming SN~2012dn reached {\it b}-band maximum light on 2012~July UT 24.8 \citep{Brown14}, optical spectra of SN~2012dn were obtained from day $-$14.6 to day $+$10.2, with a follow-up spectrum taken at $+$114.2 (see Table~2). In total we collected 15 spectra over the course of its first 25 days since discovery, with spectroscopic details probed out to late post-maximum light phases, and are shown in the left panel of Figure~\ref{Fig:specs}. Complimentary observations extend into post-maximum phases by \citet{Chakradhari14} who collected ten additional spectra from day $+$10.4 to day $+$98.2, in addition to six photospheric phase spectra. 

\section{Light Curves}

\subsection{Absolute Magnitudes}

To estimate a date of maximum light, we utilize the multicolor light-curve fitter, \texttt{MLCS2k2} \citep{Jha07}, with our fit shown in the right panel of Figure~\ref{Fig:phot}.\footnote{\url{http://www.physics.rutgers.edu/~saurabh/MLCS2k2}/} We find that SN~2012dn reached {\it B}-band maximum light on MJD 56132.8 (2012 July UT 24.8), which is consistent with that found from {\it Swift}/UVOT broadband photometry obtained by \citet{Brown14}. 

Assuming $A_{r}^{MWG}$~=~0.14 mag \citep{Schlafly11} and a Tully-Fisher distance modulus $\mu$~=~33.15~$\pm$~0.52 \citep{Springob07}, an observed peak {\it r}-band magnitude of 14.26 implies M$_{r}^{peak}$~$\approx$~$-$19.03. (\citealt{Brown14} assume $\mu$~=~33.32~$\pm$~0.20 and \citealt{Chakradhari14} use 33.15~$\pm$~0.15.) For {\it g}-band estimates, we find M$_{g}^{peak}$~=~$-$19.09 for $A_{g}^{MWG}$~=~0.20 mag.

To estimate the intervening host galaxy extinction, we measure and find an upper limit on the equivalent width of a narrow Na~D doublet to be $\lesssim$ 0.30~\AA. Based on a relation from \citet{Poznanski12}, we estimate the reddening due to the host, E(B$-$V)$_{host}$, to be $\lesssim$ 0.03 mag. By contrast, \citet{Chakradhari14} report a color excess of E(B$-$V)$_{host}$ = 0.12 mag from reportedly resolved Na~D doublet measurements of 0.69 and 0.78, respectively. This implies $A_{B}^{host}$~=~0.75 mag and M$_{B}^{peak}$~=~$-$19.52, where uncertainties may be imparted by moderately low-resolution optical spectra. Here we assume M$_{B}^{peak}$~$\lesssim$~$-$19.0 (m$_{{\it B}}$~$-$~$\mu$~$-$~A$_{B}^{MWG}$). In the case of moderate extinction, $A_{B}^{host}$$\sim$~0.4, it follows that M$_{B}^{peak}$~$\lesssim$~$-$19.4. 

We look to compare this with our light curve fit in Figure~\ref{Fig:phot}, where \texttt{MLCS2k2} handles reddening of the intrinsic spectral energy distribution through two parameters, $\Delta$ and A$_{V}$; $\Delta$ is functionally dependent on the color difference between high and low stretch light curves such that a SN~Ia with $\Delta$~=~1 (and A$_{V}$~=~0) is redder than in the case of $\Delta$~=~0. For SN~2012dn, \texttt{MLCS2k2} returns $\Delta$~=~$-$0.56~$\pm$~0.02, which is out of range for the model, but is also indicative of an intrinsically blue event. 

To match the observed colors, \texttt{MLCS2k2} compensates for $\Delta$~=~$-$0.56 by overestimating additional reddening to A$_{V}$~=~0.69~$\pm$0.03; because the shape of SN~2012dn's light curves are not normal, \texttt{MLCS2k2} returns an overestimate of host extinction, A$_{V}^{Host}$~$\sim$~0.52. If there were half a magnitude of extinction from the host galaxy of SN~2012dn, one might also expect to have a redder {\it g}~$-$~{\it i} color at maximum light. Instead, below in \S3.2 we find the {\it g}~$-$~{\it i} color of SN~2012dn is bluer than for SN~2006gz and slightly redder than SN~2009dc during the epoch of peak brightness. 

We also infer values for host-extinction with the \texttt{SALT2.4} light curve fitter \citep{Guy07}, and we find them to be roughly consistent with those obtained above with \texttt{MLCS2k2}.\footnote{\url{http://supernovae.in2p3.fr/salt/doku.php}} \texttt{SALT2} returns an apparent {\it B}-band magnitude of m$_{B}^{peak}$~=~14.25~$\pm$~0.03. The values obtained for the c parameter (0.24~$\pm$~0.03), which measures the correlation between (intrinsic plus reddened) color and peak brightness, and the x$_{1}$ parameter (1.81~$\pm$~0.07), which describes the correlation between light curve width and peak brightness, are indicative of A$_{V}$~$\sim$~0.6 and \texttt{MLCS2k2} $\Delta$~$\sim$~$-$0.4 (cf. \citealt{Kessler09}). 

An additional uncertainty in the magnitude estimate for SN~2012dn is from the distance modulus, where SN~2012dn could have reached a peak luminosity of $M_{g}$~$\approx$~$-$19.67 ($M_{B}$~$\approx$~$-$19.69 in {\it B}-band; \citealt{Chakradhari14}). However, even this approximate upper limit on the inferred peak luminosity for SN~2012dn is not far from the luminous end of seemingly spectroscopically ``normal'' events such as SN~1995D (M$_{B}^{peak}$~$\approx$~$-$19.66; \citealt{Contardo00}) and 1999ee (M$_{B}^{peak}$~$\approx$~$-$19.85; \citealt{Stritzinger02}). 

Unlike \texttt{MLCS2k2}, \texttt{SALT2} does not return an explicit distance. Therefore, it must be derived from calibration that depends on cosmological parameters. To test systematics, we adopt the two calibrations of \citet{Kessler09} and \citet{Guy10} of $\mu$~=~32.987~$\pm$~0.112 and 32.813~$\pm$~0.145, respectively, to obtain $\bar{\mu}$~=~32.9~$\pm$~0.1, which is much lower than $\mu$~=~33.24 from \texttt{MLCS2k2}. If the distance to ESO~462-16 is indeed shorter, this still implies a peak brightness for SN~2012dn that is lower than expected for its spectroscopic class. Without accounting for reddening from the host, the difference between peak absolute magnitudes inferred from \texttt{MLCS2k2} and \texttt{SALT2} is $\sim$~0.02 magnitudes in the {\it B}-band. 

In spite of the uncertain host-galaxy extinction estimate for SN~Ia in general (\citealt{Phillips13,Foley14}; see also \citealt{Cikota16}), our observations indicate M$_{g}^{peak}$ for SN~2012dn is not exceptionally luminous, and is in agreement with previous observations of \citet{Brown14} and \citet{Chakradhari14}. In fact, SN~2012dn's M$_{g}^{peak}$ is possibly below that for the Core Normal SN~1990N (see Figure~\ref{Fig:phot}). 

Assuming negligible host extinction and M$_{{\it B}}^{peak}$~=~$-$18.77, the lower limit on the peak luminosity would suggest a relatively broad range of intrinsic luminosities for this spectroscopic subclass and its progenitor systems \citep{Maeda09,Scalzo10,Brown14}. Yet by day~$+$330 post-maximum light, SN~2009dc and SN~2012dn decline to absolute {\it B}-band luminosities that are approximately half a magnitude greater than SN~2011fe. (M$_{B}^{09dc}$~$\sim$~$-$12.02~$\pm$~0.19, \citealt{Taubenberger11}; M$_{B}^{12dn}$~$\sim$~$-$12.39~$\pm$~0.66, this work and \citealt{Chakradhari14}; M$_{B}^{11fe}$~$\sim$~$-$11.57~$\pm$~0.19, \citealt{Munari13}.)

Whether or not its original system was in excess of M$_{Ch}$, there are other photometric characteristics to suggest that SN~2012dn is a peculiar member of the SCC SN~Ia subclass. Namely, the 1.08~mag decline in the {\it B}-band 15~days after peak brightness is only slightly less than those for the brighter end of normal SN~Ia, e.g., SN~1990N, 1998bu, 2003du, all of which show signatures of \ion{C}{II} in the outermost layers. In terms of the empirical light curve method of \citet{Conley08}, \texttt{SiFTO} fits to a light curve stretch, s~=~1.28, for SN~2012dn {\it g} and {\it r}-band light curves. For comparison, s~$\sim$~0.81 for SN~2006D; 0.82, 1994D; 0.86, 1996X; 0.91, 2002bo; 0.93, 2002dj; 0.96, 2005cf; 1.03, 2011fe; 1.04, 2003fg; 1.11, 1999aa; 1.26, 2001ay; and s~$\approx$~1.29 for 2009dc.

An additional characteristic of SN~2012dn that indicates a non-standard, perhaps SN~2009dc-like origin is the lack of well-pronounced secondary {\it i}-band maximum, or less than what is typically observed for spectroscopically normal SN~Ia \citep{Chakradhari14}. Within the context of the width-luminosity relationship \citep{Khokhlov93,Phillips93,Phillips99}, the suppression of a secondary maximum is consistent with spectroscopic indications that a relatively late transition from doubly to singly-ionized iron-peak elements has occurred (\citealt{Kasen07}, and see our \S5). If there is a less pronounced secondary {\it i}-band maximum for SN~2012dn, it would have occurred $\sim$10 days or more ahead of ``normal'' SN~Ia.

\begin{figure}
\centering
\includegraphics[scale=0.4]{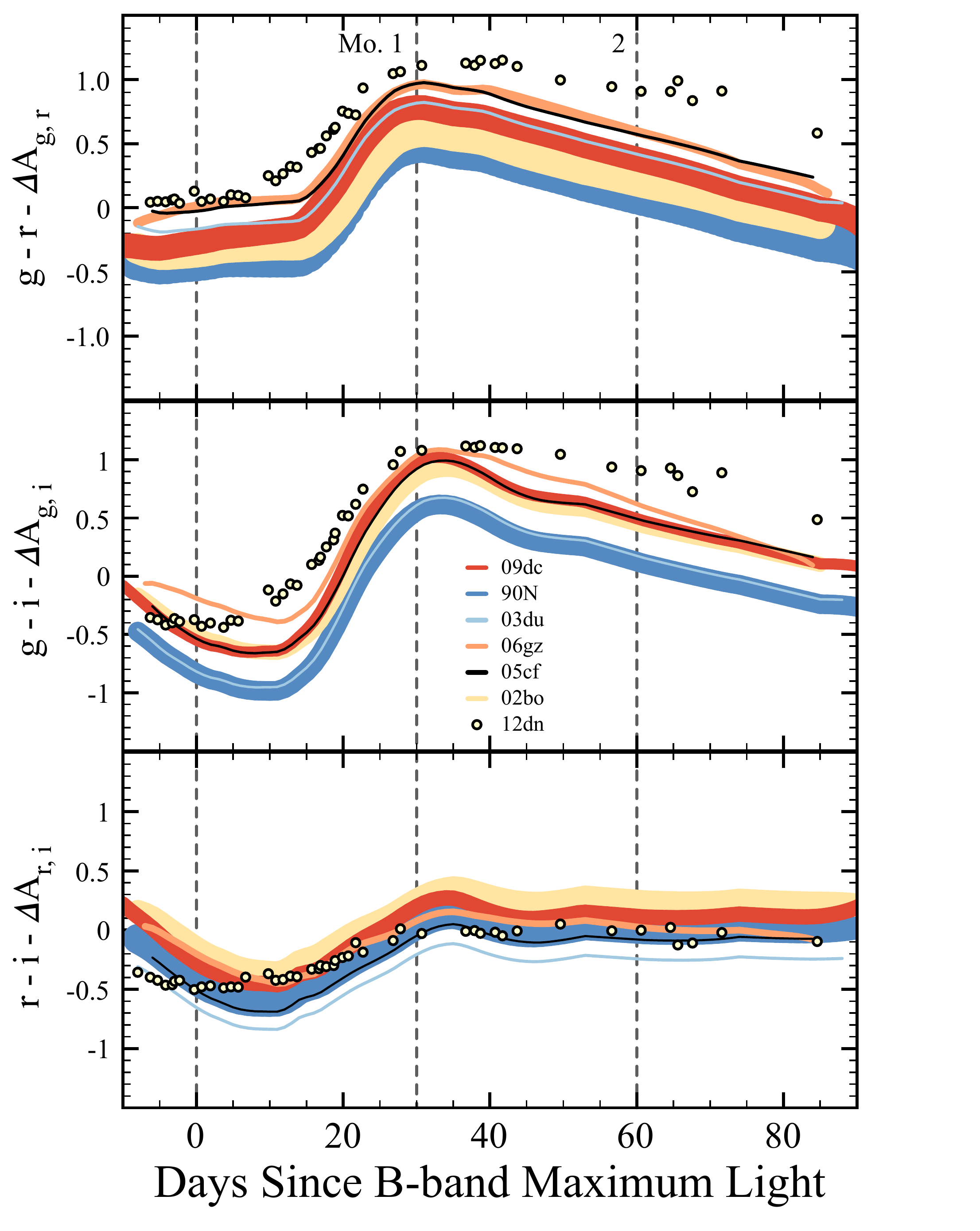}
\caption{Plotted top-down are {\it g}$-${\it r}, {\it g}$-${\it i}, and {\it r}$-${\it i} photometric color comparisons between SN~2012dn (yellow points) and super-Chandrasekhar candidate SN~2006gz and 2009dc, Core Normal SN~1990N, 2003du, and 2005cf, and the Broad Line SN~Ia, SN~2002bo \citep{Ruiz92,Phillips92,Benetti04,Anumpama05,Stanishev07,Pastorello07b,WangX09,Scalzo10}. Individual color-curves have not been corrected for reddening from the host-galaxy.}
\label{Fig:photcolors}
\end{figure}

\subsection{Colors}

The UV flux of SN~Ia is wildly variable compared to optical counterparts, which has highlighted UV wavelength regions as a capable, albeit challenging probe for physics unseen by optical wavelengths \citep{Kirshner93,WangX12,Milne13,Mazzali14}. Compared to SN~2009dc, SN~2005cf and other normal SN~Ia are typically redder at NUV wavelengths prior to maximum light \citep{Brown09,Milne10,Brown10,Brown12}, with some fraction of ``carbon-positive'' events being relatively bluer at NUV wavelengths \citep{Thomas11,Milne13}. \citet{Brown14} report the early UV colors of SN~2012dn-like events are bluer than normal as well. 

Plotted in Figure~\ref{Fig:photcolors} are our {\it g}$-${\it r}, {\it g}$-${\it i}, and {\it r}$-${\it i} colors for SN~2012dn, select SCC SN~Ia, and several spectroscopically normal SN~Ia for reference. Without correcting for dust extinction internal to the host galaxy, the evolution of SN~2012dn's {\it g}$-${\it r} color-curve during the first month post-explosion is moderately comparable to the Core Normal SN~2005cf (M$_{V}^{peak}$~=~$-$19.30~$\pm$~0.33) and a SCC, SN~2006gz (M$_{V}^{peak}$~=~$-$19.56~$\pm$~0.18). The overall {\it g}$-${\it r} color of SN~2012dn also appears redder than other SN~Ia in Figure~\ref{Fig:photcolors}. Perhaps this is either indicative of significant host-galaxy extinction, which is counter to what we find through equivalent-widths of narrow lines of Na~D (\S3.1), or the overall color-evolution might suggest a dusty environment local to SN~2012dn \citep{Maeda09,Taubenberger13}. SN~2005cf and 2006gz are also $\sim$~30\% redder than the significantly brighter SN~2009dc (M$_{V}^{peak}$~=~$-$~20.89~$\pm$~0.54) on this scale.

We attribute the early departure of SN~2012dn's {\it g}$-${\it i} and {\it r}$-${\it i} colors toward redder values during early post-maximum phases to a relatively slow-declining {\it i}-band light curve of 0.01~mag~day$^{-1}$ compared to 0.06~mag~day$^{-1}$ for {\it g}-band. Between day $+$30 and day $+$85, SN~2012dn's {\it g}-band light curve declines by 0.03~mag~day$^{-1}$. SN~2007if underwent a similar decline of ~$\sim$~0.02~mag~day$^{-1}$ on similar timescales \citep{Scalzo10}. Compared to Core Normal SN~Ia where the decline in {\it g}-band is also $\sim$~0.03~mag~day$^{-1}$, a relatively hastened late-time decline in brightness is only observed for SN~2012dn at NUV and MUV wavelengths \citep{Brown14,Chakradhari14}. 

\section{Comparison of Spectra}

\begin{figure*}
\centering
\includegraphics[scale=0.6]{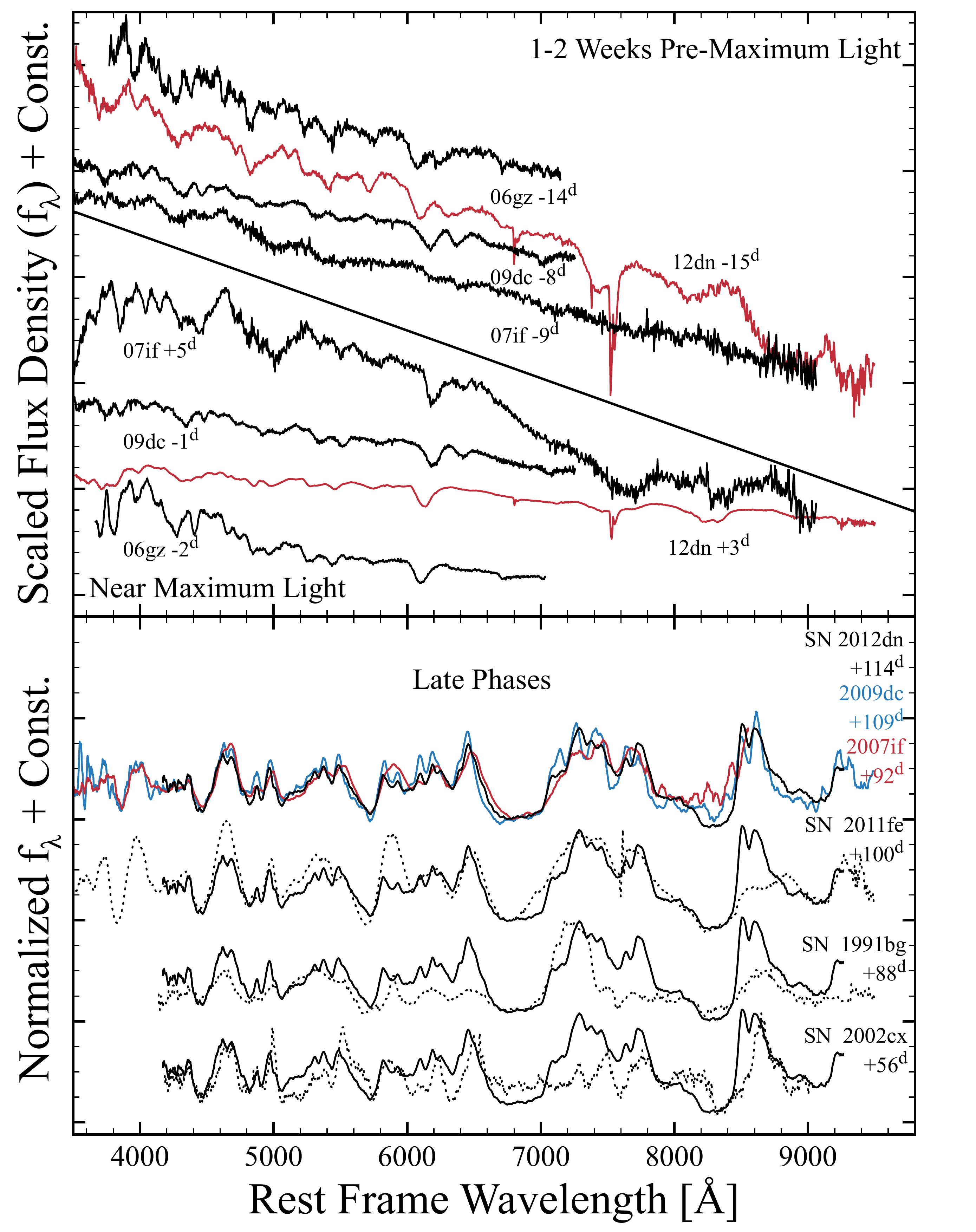}
\caption{{\it Top}: Optical spectroscopic comparisons to three super-M$_{Ch}$ candidate SN~Ia, 2006gz, 2007if, and 2009dc, during the earliest and near maximum light epochs. {\it Bottom}: Late post-maximum light phase comparisons of SN~1991bg, 2002cx, 2007if, 2009dc, 2011fe, and 2012dn; SN~1991bg, 2002cx, and 2011fe are the dotted lines.}
\label{Fig:comp}
\end{figure*}

To assess the early spectroscopic evolution and approximate overlap of ejected carbon, oxygen, and select intermediate-mass elements, in the right panel of Figure~\ref{Fig:specs} we plot the photospheric phase spectra of SN~2012dn in terms of line velocities $\mathcal{O}$(v/c) for \ion{Si}{II}~$\lambda$6355, the \ion{Ca}{II} infrared triplet weighted by oscillator strengths ($\lambda$8579), \ion{C}{II}~$\lambda$6580, and \ion{O}{I}~$\lambda$7774. For SN~2012dn we find that faster unburned and Si-rich material overlap, in part, below the outer extent of Ca-rich absorbing layers, which is typical for Core Normal SN~Ia in terms of singular line velocities. See \S5 where our \texttt{SYNAPPS} analysis includes the relativistic correction term for estimating projected Doppler velocities.

\subsection{Photospheric Phases}


In the top panel of Figure~\ref{Fig:comp}, comparisons of SN~2012dn to SCCs SN~2006gz, 2007if, and 2009dc during photospheric epochs reveal moderate similarities across optical wavelengths. All share a fairly comparable pseudo-continuum level in addition to superimposed atomic signatures during this phase. Based on dissimilar appearances at blue-ward wavelengths, one could argue SN~2007if, 2009dc, and 2012dn do not belong to a similar class of events, particularly since too few events have been studied to know which if any is the most normal prototype. Still, while visual comparisons between SN~2012dn and other SCC SN~Ia during photospheric phases suggest a match for this subtype, accurate classifications of SN benefit substantially from multi-epoch observations. 
 
\subsection{Later Phases}

In the bottom panel of Figure~\ref{Fig:comp}, we compare late post-maximum spectra of SN~2012dn to a variety of peculiar events. The spectra of SCC SN~Ia and SN~2012dn share fairly similar characteristics apart from small line shifts in individual features. It is thus reasonable to suspect that SN~2007if, 2009dc, and 2012dn are of a similar spectroscopic subclass \citep{Brown14,Chakradhari14}.

The moderate contrast between late-time spectra of SN~2012dn, a Core Normal (SN 2011fe), Cool (SN 1991bg), and a more peculiar SN~Ia (SN 2002cx; see \citealt{McCully14b,McCully14a,Foley1408ha,Stritzinger15}) suggests SN~2012dn is not strictly ``well-matched'' with normal nor historically peculiar SN~Ia. In particular, the 4700 and 5900~\AA\ emission line regions of SCC SN~Ia are not as prominent compared to that of SN~2011fe, possibly due to relatively sustained heating or less nickel synthesized during the explosion. This might make sense given that SN~1991bg and 2002cx, as sub-luminous SN~Ia, also show less pronounced 4700~\AA\ and 5900~\AA\ features.

\section{Semi-empirical line identifications}

For SN spectral features the accuracy of empirical measurements depends on the prescriptions of lines utilized from a supporting abundance model \citep{HatanoAtlas,Branch07a}. Studies of SN diversity thus benefit when prominent detections of select ions can be ruled out. Considering the homogeneous nature of SN~Ia spectra, the narrow features of SCC SN~Ia lend themselves particularly useful for scrutinizing semi-empirical line-identification schema (e.g., \citealt{Scalzo10,Doull11}, and see also \citealt{Foley10a}).

\subsection{The \texttt{SYNAPPS} Model}

To estimate line velocities of select atomic species, we utilize the fast and highly parameterized spectrum synthesis code, \texttt{SYNAPPS} \citep{ThomasSYNAPPS}. The \texttt{SYNAPPS} model, formerly \texttt{SYNOW} \citep{Fisher00}, assumes spherical symmetry, a sharp photosphere, pure resonance line scattering treated under the Sobolev approximation, and Boltzmann statistics for electron level populations (parameterized by \texttt{temp} in \texttt{SYNAPPS}). For a given reference line, the radial dependence of optical depth is assumed to be an outwardly decreasing exponential that is dependent on the optical depth at the photosphere (\texttt{logtau}), the photospheric velocity (\texttt{vphot}), the detachment velocity for an ion (\texttt{vmin}), and an e-folding parameter (\texttt{aux}). Optical depths of the remaining lines assume LTE level populations.

\texttt{SYNAPPS} essentially views photospheric phase supernova spectra as a non-linear superposition of resonance line P~Cygni profiles atop an underlying pseudo-continuum level. While spectrum formation is not similarly trivial for supernova atmospheres \citep{Bongard08,Friesen14}, the one-dimensional \texttt{SYNAPPS} model has served as a complimentary tool to direct comparisons of observed spectra. In certain situations, the rest-wavelength emission component can undergo a similar blue-shift, as seen for the absorption component \citep{Baron93,Anderson14}. As the ejecta expands, geometric dilution may cause the rest-wavelength emission to undergo a red-shift during later post-maximum epochs \citep{Friesen12}. The \texttt{SYNOW/SYNAPPS} model is currently unable to account for these effects on account of an assumed sharp photosphere.

Two of the top-most \texttt{SYNAPPS} parameters are the relative velocities of included ions, \texttt{v$_{min}$}, and the relative strengths of the lines which can be set by \texttt{log~$\tau$}. By hypothesizing the prescription of lines that influence each composite spectral feature, a synthetic spectrum can be minimized to the data for the available parameter space. For \texttt{SYNAPPS}, this is most often \texttt{log~$\tau$}, \texttt{v$_{phot}$}, \texttt{v$_{min}$}, \texttt{aux}, and multiplied by the number of ions included. In turn, this enables a tracing of a monotonically decreasing v$_{line}$ over time ($\pm$~$\delta$$v$; resolutions set by local line-blending).
 
A reference continuum level is set by an assumed blackbody (\texttt{T$_{BB}$} in \texttt{SYNOW}; \texttt{t$_{phot}$} in \texttt{SYNAPPS}). Spectral energy distributions of SN~I are not well-represented by blackbodies \citep{Bongard06,Bongard08} and the same can be said for SN~II \citep{Hershkowitz86approx,Hershkowitz86numerical,Hershkowitz87}. However, the assumption for \texttt{SYNAPPS} is one of practicality and is not meant to be a reliable indicator of a real temperature structure (see Fig.~5 of \citealt{Bongard08}). In \texttt{SYNAPPS}, quadratic warping constants \texttt{a$_{0}$}, \texttt{a$_{1}$}, and \texttt{a$_{2}$} supplement \texttt{t$_{phot}$} to form a more flexible continuum reference level; \texttt{a$_{0}$}, \texttt{a$_{1}$}, \texttt{a$_{2}$}, and \texttt{t$_{phot}$} form the backbone to data-driven minimization. (See \citealt{ThomasSYNAPPS} and Fig.~2 of \citealt{Parrent12} for an example.)

Thus, for well-observed events, one can conduct various experiments to directly assess: (i)~detectable and inferable lines of noteworthy species, including C, O, Mg, Si, S, Ca, Fe, and Co at photospheric (PV) and higher velocities (HV); (ii) how to best utilize and constrain semi-empirical line-measurements; (iii) the overlap of burned and remaining progenitor material in terms of line velocities; (iv) the emergence, cutoff, and detachment velocities of select species; (v) the time-dependent prescription and related uncertainties for several composite features \citep{Scalzo14,Milisavljevic15}; and (vi) line-velocity gradients and plateaus.
 
\subsection{Application of \texttt{SYNAPPS}}

Representative ``best fits'' to spectra using \texttt{SYNAPPS} are typically initialized by exploring single and multi-ion comparisons while perturbing the available parameters. As mentioned above, because parameterized expansion velocities, $v_{line}$, are traced by blended atomic signatures over time, the convergence of \texttt{v$_{min,Y}$}(t) with \texttt{SYNAPPS} depends on the list of actively contributing atomic species used as input. However, since \texttt{SYNAPPS} does not solve for ionization balance, which is a necessary step for abundance determinations and more detailed analysis \citep{Baron96}, semi-empirical fits are devised by perturbing compositional constructs \citep{HatanoAtlas}. Synthetic fits for a given set of ions can then be processed forward and backward in phase space relative to maximum light.

The mode of interpretation utilized by \texttt{SYNAPPS} is one of an onion shell paradigm for the radiation transport in supernova atmospheres, which is innately limited (cf. \citealt{Blondin15}). However, the \texttt{SYNOW} model can be used to constrain the compositional structure of ejected material in spite of being confined to Saha-Boltzmann statistics \citep{Branch07a,Parrent10}. A standard prescription of ions for SN~Ia stems from \citet{Hatano99,HatanoAtlas} and \citet{Branch05}, which is based on the time-dependent inclusion of ions including \ion{C}{II}, \ion{O}{I}, \ion{Mg}{II}, \ion{Si}{II}, \ion{Si}{III}, \ion{S}{II}, \ion{Ca}{II}, \ion{Cr}{II}, \ion{Fe}{II}, \ion{Fe}{III}, and \ion{Co}{II}.

Previously, \citet{Branch77} showed how the location of an absorption minimum does not always correspond to the model photospheric velocity. This was found to occur for both weak and saturated lines. To ensure minimal under and over-shooting in $v_{min}$ for each spectrum of SN~2012dn (and similarly for SN~2011fe in \citealt{Parrent12}), we re-examined our preliminary results for a range of saturated optical depths for each ion. Additionally, when $v_{min}$ for any ion was found to be off by more than 1000~km~s$^{-1}$ (a common occurrence for \ion{Fe}{II}, \ion{Fe}{III}, and HV~signatures), we perturbed $v_{min}$ by $\pm$~1000~km~s$^{-1}$, and repeated the above fitting methodology for the entire dataset. In the event that the fitting procedure produced either monotonically increasing $v_{min, Y}$ over time, or oscillatory solutions with amplitudes exceeding 1000~km~s$^{-1}$, we interpreted this as an artifact of adjacent and unaccounted for lines, $v_{min, Z}(t)$, rather than blue-ward shifting absorption signatures during a period of photospheric recession \citep{Branch73,Kirshner73}.

Because we are interested in extracting best estimates of $v_{min}$ for various species, we initially fixed \texttt{aux} and \texttt{temp} at default values of 10$^{3}$~km~s$^{-1}$ and 10$^{4}$~K, respectively, for all but \ion{Ca}{II} where we allowed \texttt{aux} to take on values greater than 1000~km~s$^{-1}$ (see \citealt{Branch05}). These parameters were then free to vary near the end of our fitting procedure. We also allowed \texttt{logmin} to vary freely within appropriate ranges when minimizing $v_{min, Y}$ toward a monotonically decreasing function of time (see \citealt{HatanoAtlas,Branch05}).

Hence, we define ``best fits'' as satisfying the condition that a fit starting with iron-peak elements converge to a similar consensus in $v_{min, Y}$ for a fit ending with the inclusion of iron-peak elements (i.e., $v_{min}$ to within 1000~km~s$^{-1}$ for each ion considered). With approximately three to five seconds per output spectrum on average, the above vetting process yields on the order of 1~$-$~2~x~10$^{3}$ synthetic spectrum comparisons before $v_{min,Y}(t)$ begins to converge toward a global minimum in the available parameter space.

\section{Spectral Analysis and Results}

\begin{figure}
\centering
\includegraphics*[scale=0.33, trim = 10mm 0mm 0mm 0mm]{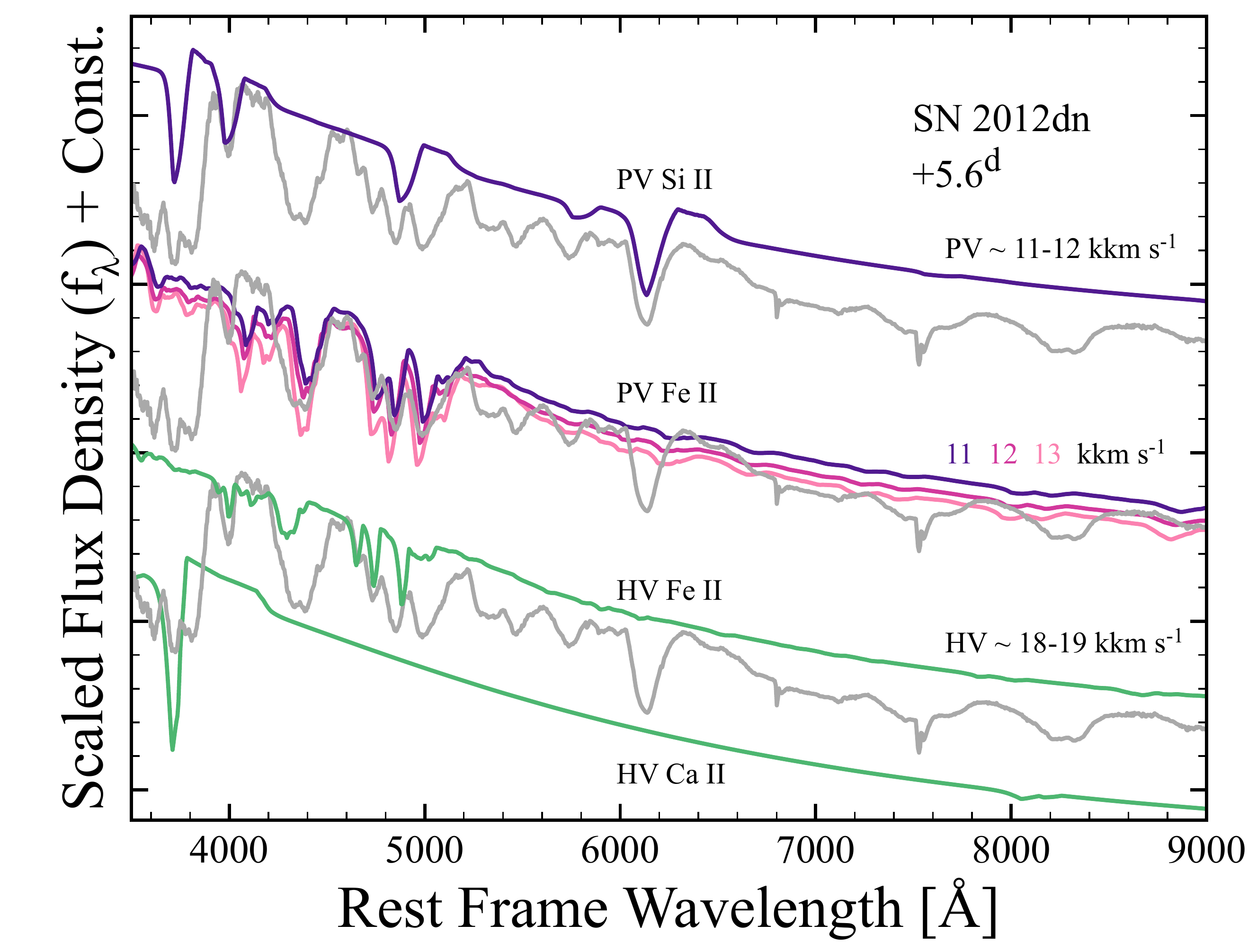}
\caption{Plotted alongside the day $+$5.6 spectrum of SN~2012dn (in grey) are single-ion \texttt{SYN++} synthetic spectra for photospheric \ion{Si}{II} and \ion{Fe}{II}, in addition to detached, high velocity \ion{Ca}{II} and \ion{Fe}{II}. We have exaggerated the strength of HV~\ion{Ca}{II} lines with \texttt{SYN++} to show the weak infrared triplet signature. See text in \S6.2. Model \texttt{v$_{min}$} are given in units of km~s$^{-1}$~1000$^{-1}$. }
\label{Fig:comps_0730}
\end{figure}

In this section we discuss a breakdown of our results, which are shown in Figures~\ref{Fig:comps_0730}$-$\ref{Fig:carboncomps} and later Figure~\ref{Fig:paradisefig} in \S7.

\subsection{Photospheric Velocities of \ion{C}{II}}

\begin{figure}
\centering
\includegraphics*[scale=0.33, trim = 10mm 0mm 0mm 0mm]{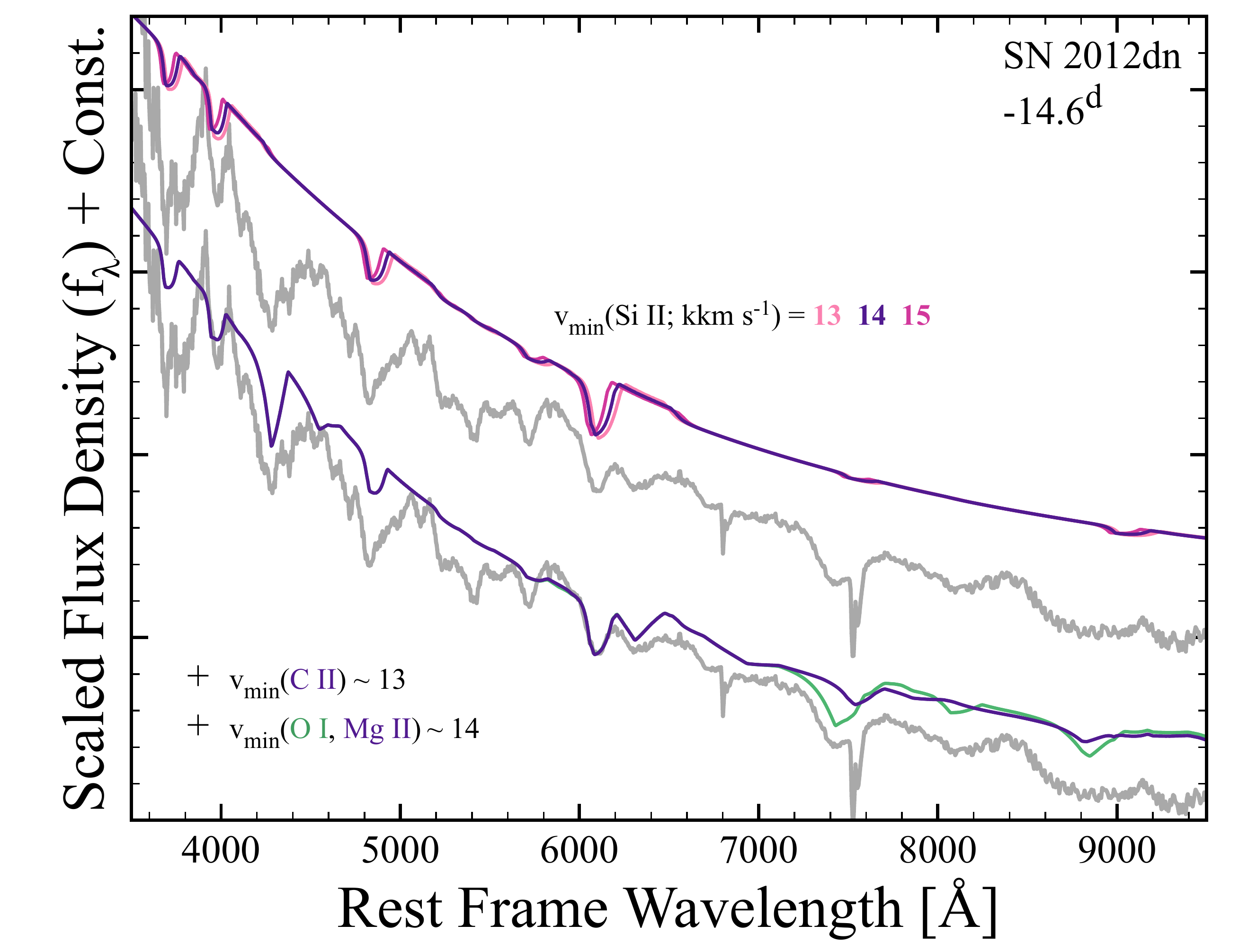}
\caption{Representative composite \texttt{SYN++} spectra for \ion{Si}{II}, \ion{C}{II}$+$\ion{Mg}{II}$+$\ion{Si}{II}, and a single-ion spectrum of \ion{O}{I} compared to the day $-$14.6 of SN~2012dn (in grey). Flux scale is arbitrary and model \texttt{v$_{min}$} are given in units of km~s$^{-1}$~1000$^{-1}$. Unlike for gaussian fitting methods and pseudo-equivalent widths of blended spectral features, the plus symbols here and for the remaining figures do not imply that line formation for \texttt{SYNAPPS} is treated as linearly additive.} 
\label{Fig:comps_0710}
\end{figure}

Owing in part to a lack of strong evidence for signatures of high velocity features compared to more normal SN~Ia, the approximate rate of change in line velocities is reportedly lower than normal for SCC SN~Ia. Nonetheless, both SN~2006gz and 2012dn have photospheric velocities typical of normal SN~Ia despite narrower than normal spectral features (see Fig.~20 of \citealt{Chakradhari14}). During early photospheric phases, our analysis indicates line velocities of $\sim$~14,000~km~s$^{-1}$ for \ion{Si}{II}. From day $-$14.6 to day $+$5.6, line velocities are consistent with having only reduced by 2000~$-$~3000~km~s$^{-1}$ to 11,000~$-$~12,000~km~s$^{-1}$ (cf. Figure~\ref{Fig:comps_0730}~and~\ref{Fig:comps_0710}).

Representative single-ion spectra of \ion{C}{II}, \ion{O}{I}, and \ion{Si}{II} are compared to SN~2012dn on day $-$14.6 in Figure~\ref{Fig:comps_0710}. For \ion{C}{II}, the value of \texttt{v$_{min}$} that our full time-series fitting converges towards is slightly below that estimated for \ion{Si}{II} (12,000~$-$~13,000~km~s$^{-1}$ and 14,000~$-$~15,000~km~s$^{-1}$, respectively) and is outside typical resolutions for \texttt{v$_{min}$} of $\sim$~500~$-$~1000~km~s$^{-1}$; i.e., we confirm that an improved estimate of \texttt{v$_{min}$} for \ion{C}{II} is not precisely 14,000~km~s$^{-1}$, which is consistent with an independent \texttt{SYNOW} analysis by \citet{Chakradhari14}. If such low projected Doppler velocities reflect ejecta asymmetries, the slight mismatch with \texttt{v$_{min}$} for \ion{Si}{II} could be due to an off-center distribution of carbon.

Finally we caution that lines of \ion{Fe}{II} and \ion{Co}{II} are available for contaminating weak 6300~\AA\ features by early post-maximum light phases, thereby further motivating the tracing of time-varying inferences of \ion{Fe}{II}. However, by estimating P~Cygni summation of \ion{Cr}{II} and \ion{Co}{II} in the NUV \citep{Scalzo10,Scalzo14,Brown14}, it is possible to provide a soft upper limit to the blended strength of \ion{Fe}{II} and \ion{Co}{II} that are permissible near 6300~\AA\ (see \S6.2 below).

\subsection{Direct Inference of Detached Species}

From the epoch of maximum light onward, lines of \ion{Fe}{II} are expected to influence a number of features at optical wavelengths (cf. Figure~\ref{Fig:comps_0730} and Fig.~2 of \citealt{Parrent12}). However, while small contribution from \ion{Fe}{II} can be inferred near day $+$5.6, \ion{Fe}{II} has yet to create a conspicuous and therefore traceable signature during earlier phases. This interpretation is marginally consistent with the lack of a well-pronounced i-band double peak for SN~2012dn \citep{Kasen06NIR,Chakradhari14}. 

\begin{figure}
\centering
\includegraphics*[scale=0.35]{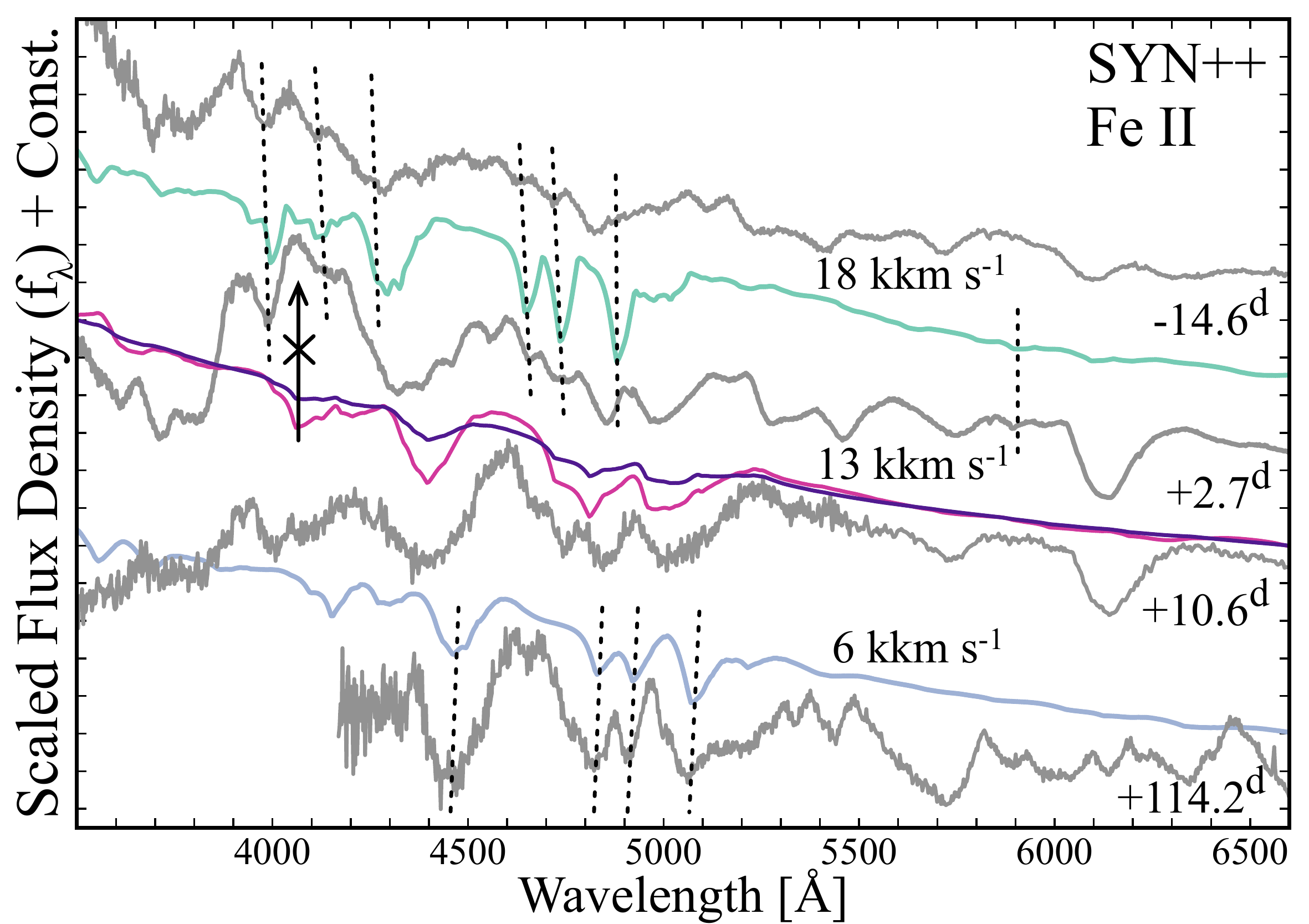}
\caption{\texttt{SYN++} single ion comparison for \ion{Fe}{II} at 6000, 13,000, and 18,000~km~s$^{-1}$. At post-maximum light phases (after day $+$10.6), \ion{Fe}{II} overlaps nicely with the observed spectrum as a species formed near the photospheric line forming region. At early epochs (day $-$14.6), high velocity \ion{Fe}{II} is a promising match. The arrow marked with $\times$ signifies that contributions from \ion{Fe}{II} are restricted by the data at this wavelength for the given \texttt{v$_{min}$}.}
\label{Fig:FeII}
\end{figure}

We interpret the lack of obvious \ion{Fe}{II} signatures from an incompatible match for the 4100~\AA\ feature between the model \ion{Fe}{II} and observations (see Figure~\ref{Fig:FeII}); i.e. prior to maximum light, the evolution of SN~2012dn's spectra do not reveal signatures of \ion{Fe}{II} that can be easily distinguished from other species like \ion{Si}{II}, \ion{Si}{III}, \ion{S}{II}, \ion{S}{III}, and \ion{Fe}{III}. 
 Still, this particular ``non-detection'' of PV \ion{Fe}{II} does not remove (PV~$+$~HV)~\ion{Fe}{II} from influencing the spectrum during this epoch. Rather, the overall structure of the spectrum is difficult to reconcile for a standard prescription of PV~\ion{Fe}{II} as prescribed with LTE \texttt{SYNAPPS}.  

Ultimately, the inference of high velocity iron-peak elements is currently best served via more detailed modeling. However, to see more clearly when and where lines of (PV~$+$~HV) \ion{Fe}{II} sync with SN~2012dn spectra, in Figure~\ref{Fig:FeII} we show synthetic single-ion spectra of \ion{Fe}{II} for \texttt{v$_{min}$} set to 18,000, 13,000 and 6000~km~s$^{-1}$ and compare them, respectively, to the day $-$14.6, near-maximum light, and late-time spectra of SN~2012dn.\footnote{See also \citet{Hatano99} and \citet{Branch05} who investigated signatures of HV \ion{Fe}{II} in the normal SN~Ia~1994D.} During early ``nebular'' phases, the presence of permitted \ion{Fe}{II} lines would be most consistent near 6000~km~s$^{-1}$.

\begin{figure}
\centering
\includegraphics*[scale=0.33, trim = 10mm 0mm 0mm 0mm]{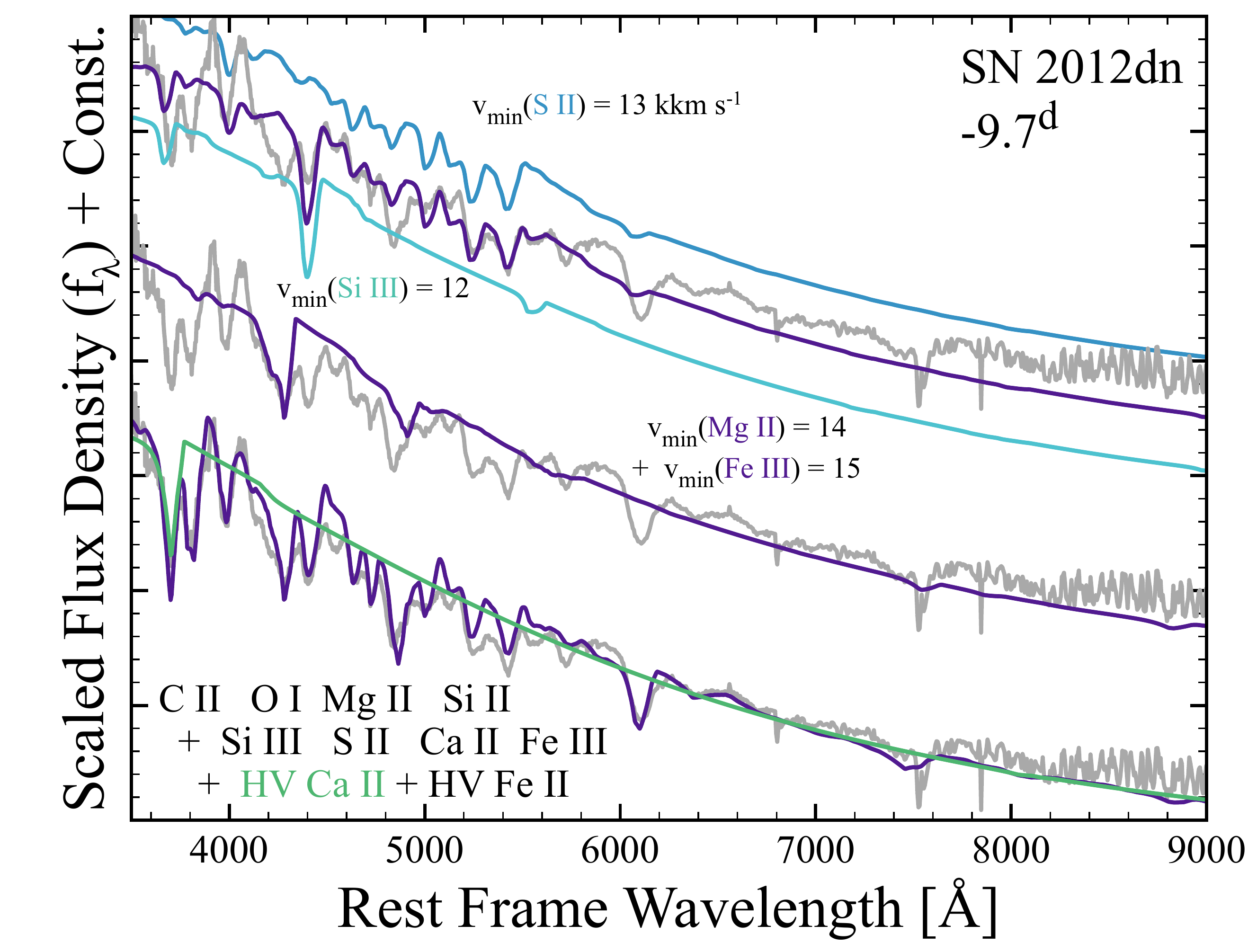}
\caption{Single-ion and full fit comparisons to the day $-$9.7 spectrum of SN~2012dn (plotted three times in grey for clarity). Model \texttt{v$_{min}$} are given in units of km~s$^{-1}$~1000$^{-1}$ and ``HV'' denotes detached species at higher than photospheric \texttt{v$_{min}$}.}
\label{Fig:comps_0715}
\end{figure}

If weak signatures of detached HV \ion{Ca}{II} and \ion{Fe}{II} are present in the spectra of SN~2012dn, then their respective signatures would be consistent with observations assuming line velocities above 18,000~km~s$^{-1}$ on day $+$5.6 and earlier (see~Figure~\ref{Fig:comps_0730}). Earlier on day~$-$9.7, the strength of a 4000~\AA\ absorption feature, which is significantly influenced by \ion{Si}{II}~$\lambda$4131, is noticeably larger than normal (see out Figure~\ref{Fig:comps_0715}). This would suggest the relative strengths of PV~\ion{Si}{II}~$\lambda\lambda$3858, 4131 could be utilized to rule out faint signatures of detached HV~\ion{Ca}{II} (see also \S5.4 of \citealt{Blondin13}).

To evaluate the potential influence of permitted lines of \ion{Fe}{II} and \ion{Co}{II} on longer-lasting 6300~\AA\ features during post-maximum epochs, in Figure~\ref{Fig:comps_0803} we examine the day $+$10.6 spectrum of SN~2012dn. Our full representative fit (in purple) also includes photospheric \ion{O}{I}, \ion{Mg}{II}, \ion{Si}{II}, and \ion{Ca}{II}. Line velocities of \ion{O}{I} are poorly constrained due to an obscuring telluric feature, however during these epochs \ion{O}{I} is expected to be near line velocities of \ion{Si}{II}. (This is also shown in Figure~\ref{Fig:comps_0710}.) By this time, permitted lines of \ion{Fe}{II} are consistent with observations for \texttt{v$_{min}$} near 11,000~km~s$^{-1}$. 

When we add \ion{Cr}{II} and \ion{Co}{II}, the fit (in orange) undergoes a modest improvement blue-ward of 4100~\AA. Without high signal-to-noise UV coverage, however, we are unable to push spectroscopic interpretations beyond generalizing the presence of iron-peak elements. Therefore, while we suspect 6300~\AA\ features are susceptible to contamination by permitted signatures of \ion{Cr}{II} and \ion{Co}{II} during post-maximum phases, we cannot precisely investigate the matter here. 

\subsection{Composite Features near 4110 and 4550~\AA}

The origins of the notches near 4110~\AA\ and 4550~\AA\ in early SN~Ia spectra have been difficult to determine on account of excessive line-blending from several candidate ions. For example, it would seem reasonable to associate both of these features with \ion{C}{II}~$\lambda$4267 and~$\lambda$4743, respectively. In Figure~\ref{Fig:comps_0710}, we show the close proximity of P Cygni profiles \ion{C}{II}~$\lambda$4267 to 4110~\AA\ and \ion{C}{II}~$\lambda$4743 to 4550 \AA. It would be advantageous if these features were signatures of unburned progenitor material, particularly when both features in question are in most SN~Ia pre-maximum spectra for most SN~Ia subtypes (see Figures~5$-$8 in \citealt{REVIEW}). 

Whatever the source during photospheric phases, identifying the 4110~\AA\ and 4550~\AA\ notches as primarily indicative of \ion{C}{II} line-strengths, however, is not an accurate interpretation, or at least not for \texttt{SYNAPPS/SYNOW}. A possible source for the confusion of 4550~\AA\ features as resulting from \ion{C}{II}~$\lambda$4743 is Fig.~1 of \citet{Hicken07} where the spectrum of SN~2006gz has been imprecisely labeled for this particular feature. Below we show that the corresponding \texttt{SYNOW} fit is not in fact able to confidently associate \ion{C}{II}~$\lambda\lambda$4267,~4743 with the 4110~\AA\ and 4550~\AA\ features.

\subsubsection{The 4550 \AA\ Feature}

At the top of Figure~\ref{Fig:comps_0715} we compare \texttt{SYNAPPS} single-ion \ion{Si}{III} and \ion{S}{II} spectra to the day $-$9.7 spectrum of SN~2012dn. Several narrow features are largely consistent with being shaped, in part, by \ion{S}{II} during these photospheric phases. 
When \texttt{v$_{min}$} and \texttt{log~$\tau$} for \ion{S}{II} are adjusted to reproduce the ``\ion{S}{II}~W'', overlap with the 4550~\AA\ feature is obtained without an {\it ad hoc} tuning of the remaining \texttt{SYNAPPS} parameters. Therefore, rather than \ion{C}{II}~$\lambda$4743, we suspect the 4550~\AA\ feature is instead at least \ion{S}{II}, with every other adjacent line influencing the blended strength of \ion{S}{II} (including \ion{C}{II}~$\lambda$4743 just barely red-ward of the 4550~\AA\ notch).

\begin{figure}
\centering
\includegraphics*[scale=0.35]{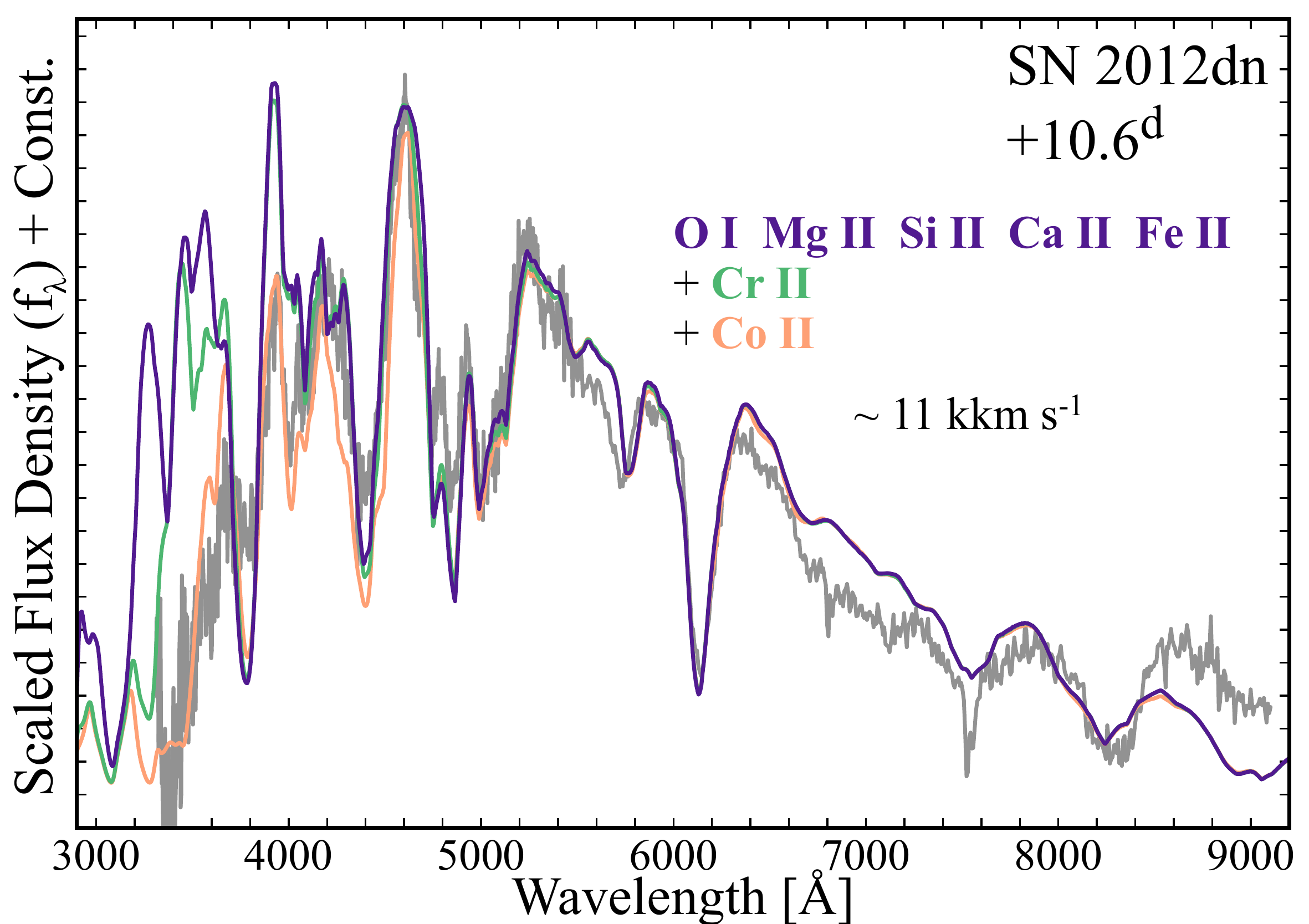}
\caption{Composite synthetic spectra compared to the day $+$10.6 spectrum of SN~2012dn (in grey). Model \texttt{v$_{min}$} are given in units of km~s$^{-1}$~1000$^{-1}$. \ion{Na}{I} has not be included in the fit, and forbidden lines of, e.g., \ion{Co}{III} are not taken into account (see \citealt{Childress15}).}
\label{Fig:comps_0803}
\end{figure}

Even when viewing the post-explosion common envelope or disk region for select super-M$_{Ch}$ merger models, the intensity minimum of a conspicuous \ion{C}{II}~$\lambda$4732 signature is assuredly red-ward of the 4550~\AA\ features, and particularly when the associated \ion{C}{II}~$\lambda$6580 is less conspicuous \citep{Raskin14}. If one were to use \texttt{SYNAPPS} and force the \ion{C}{II} to match the 4550~\AA\ feature, then \ion{C}{II}~$\lambda$6580 would not match to the 6300~\AA\ feature; i.e., the accompanying \ion{C}{II}~$\lambda$6580 signature does not hide well in the neighboring \ion{Si}{II} feature for \texttt{v$_{min}$}~=~15,000~km~s$^{-1}$ \citep{Parrent11}, nor is this scenario expected for Broad Line SN~Ia \citep{Stehle05,Blondin15}. 

One can also argue against \ion{C}{II}~$\lambda$4743 since the 4550~\AA\ feature shows up in the spectra of most SN~Ia subtypes during pre-maximum epochs. How then can near carbon-less SN~Ia (e.g., SN~2002dj, 2010jn; \citealt{Hachinger13}) still have a near-identical 4550~\AA\ feature? Instead, we suspect the feature may be primarily influenced by \ion{S}{II} in those cases as well. 

If we instead consider SN~2009dc, an extreme example where the \ion{C}{II}~$\lambda$6580 is more conspicuous than normal, why does the weak 4550~\AA\ not also follow this trend if it is attributed to \ion{C}{II}~$\lambda$4743? Our guess here would at least favor \ion{S}{II}, with possibly some contribution from \ion{C}{II} near the red-most wing of the 4550~\AA\ feature (see Figure~\ref{Fig:comps_0715}). 

From our \texttt{SYNAPPS} analysis, it is reasonable to suspect that the 4550~\AA\ region is also influenced by lines of \ion{Si}{III} and \ion{Fe}{II} that show up blue-ward of 4550~\AA\ (see Figures~\ref{Fig:comps_0710}~and~\ref{Fig:comps_0715}). That is, the stronger than normal feature immediately blue-ward of the 4550~\AA\ notch is not well-matched by \ion{C}{II}~$\lambda$4743, nor is it necessarily available. Better agreement is found with \ion{Si}{III}~$\lambda$4560, which we suspect is then influenced proportionately by \ion{Fe}{II} over time. An interpretation of \ion{Si}{III} is reasonable considering the time-dependent presence of an equally tentative detection of \ion{Si}{III} on the red-most side of the ``\ion{S}{II}~W'' (\citealt{Thomas07,Hachinger12,Sasdelli14}; and see our Figure~\ref{Fig:comps_0715} and \S6.4). This appears to be consistent with the same candidate ions from the more detailed model spectra presented by \citet{Blondin15} for the Broad Line SN~Ia 2002bo, yet additional investigations are needed to study the evolution of weak spectral signatures.

\subsubsection{The 4110 \AA\ Feature}

A similar set of conclusions follow when attempting to identify the 4110~\AA\ feature, where both \ion{C}{II} and \ion{Cr}{II} have been proposed as the primary sources:
\begin{itemize}

\item[--] Many events of varying peculiarities harbor a 4110~\AA\ feature. This feature is blended for objects with either higher mean expansion velocities or a shallower decline in opacity along the line-of-sight. By contrast, this 4110~\AA\ feature is resolved and therefore more conspicuous for objects with lower mean expansion velocities that just so happen to have conspicuous \ion{C}{II}~$\lambda$6580 absorption features, but not necessarily strong or detectable signatures of \ion{C}{II}~$\lambda$4267.

\item[--] Despite a relatively strong and long-lasting \ion{C}{II}~$\lambda$6580 absorption feature, the increasingly conspicuous 4110~\AA\ feature for SN~2007if and 2009dc disfavors the interpretation of a simultaneously prolonged detection of \ion{C}{II}~$\lambda$4267 for at least epochs nearest to maximum light and thereafter \citep{Taubenberger13}.

\item[--] In Figure~\ref{Fig:carboncomps} we show that a deep and wide \ion{C}{II}~$\lambda$7234 absorption$-$as parameterized with high Boltzmann excitation temperatures in \texttt{SYNAPPS}$-$is not comparable with the observations during the relevant epochs. Furthermore, it is likely that species including \ion{S}{II}, \ion{Cr}{II}, and (PV~$+$~HV)~\ion{Fe}{II} also influence the shape and depth of the 4110~\AA\ feature, and without invoking LTE excitation temperatures greater than 15,000~K \citep{Scalzo10,Taubenberger13}. Additionally, or at least for some Broad Lined SN~Ia during the first 22 days post-explosion, the region near 4110~\AA\ overlaps with weak signatures of \ion{S}{III}~$\lambda$4254 as well \citep{Blondin15}; i.e. this for a subtype of SN~Ia where high ionization species such as \ion{S}{III} are generally not considered for spectral decompositions. 

\begin{figure}
\centering
\includegraphics[scale=0.33, trim = 10mm 0mm 0mm 0mm]{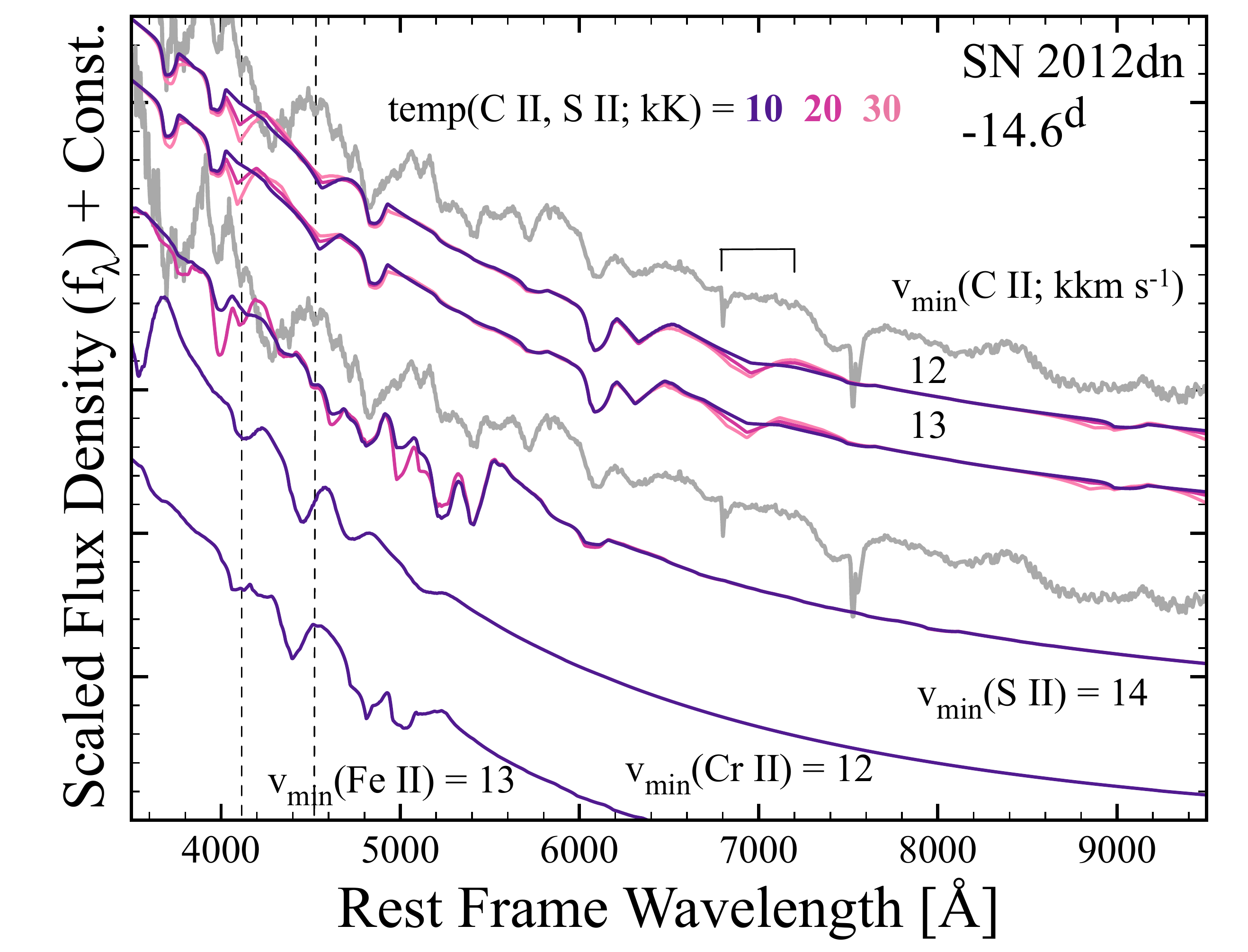}
\caption{The day $-$14.6 spectrum of SN~2012dn, plotted twice in grey for clarity, is compared to a computed spectrum including \ion{Si}{II} and \ion{C}{II}, and single-ion calculations for \ion{S}{II}, \ion{Cr}{II}, and \ion{Fe}{II}. Excitation temperatures (in units of Kelvin~x~1000$^{-1}$) and expansion velocities (in units of~km~s$^{-1}$~x~1000$^{-1}$) are varied for \ion{C}{II} and \ion{S}{II} as shown. Vertical dashed lines mark the locations of the 4100~\AA\ and 4550~\AA\ features.}
\label{Fig:carboncomps}
\end{figure}

\item[--] A \ion{C}{II}~$\lambda$4267 absorption feature is not well-matched to the 4110~\AA\ feature once the half-blended \ion{C}{II}~$\lambda$6580 line is centered near 13,000~km~s$^{-1}$, nor at 12,000~km~s$^{-1}$. If the 4110~\AA\ feature were simply due to \ion{C}{II}~$\lambda$4267 alone, then it could be used to anchor \texttt{v$_{min}$} toward slightly lower values. However, we cannot rule in \ion{C}{II}~$\lambda$4267 as primarily responsible for the shape and evolution of the 4110~\AA\ feature, and this is in agreement with notions previously expressed by \citet{Taubenberger13}.

\end{itemize}

\subsection{\ion{C}{III} $\lambda$4649}

\begin{figure*}
\centering
\includegraphics*[scale=0.48, trim = 10mm 0mm 0mm 0mm]{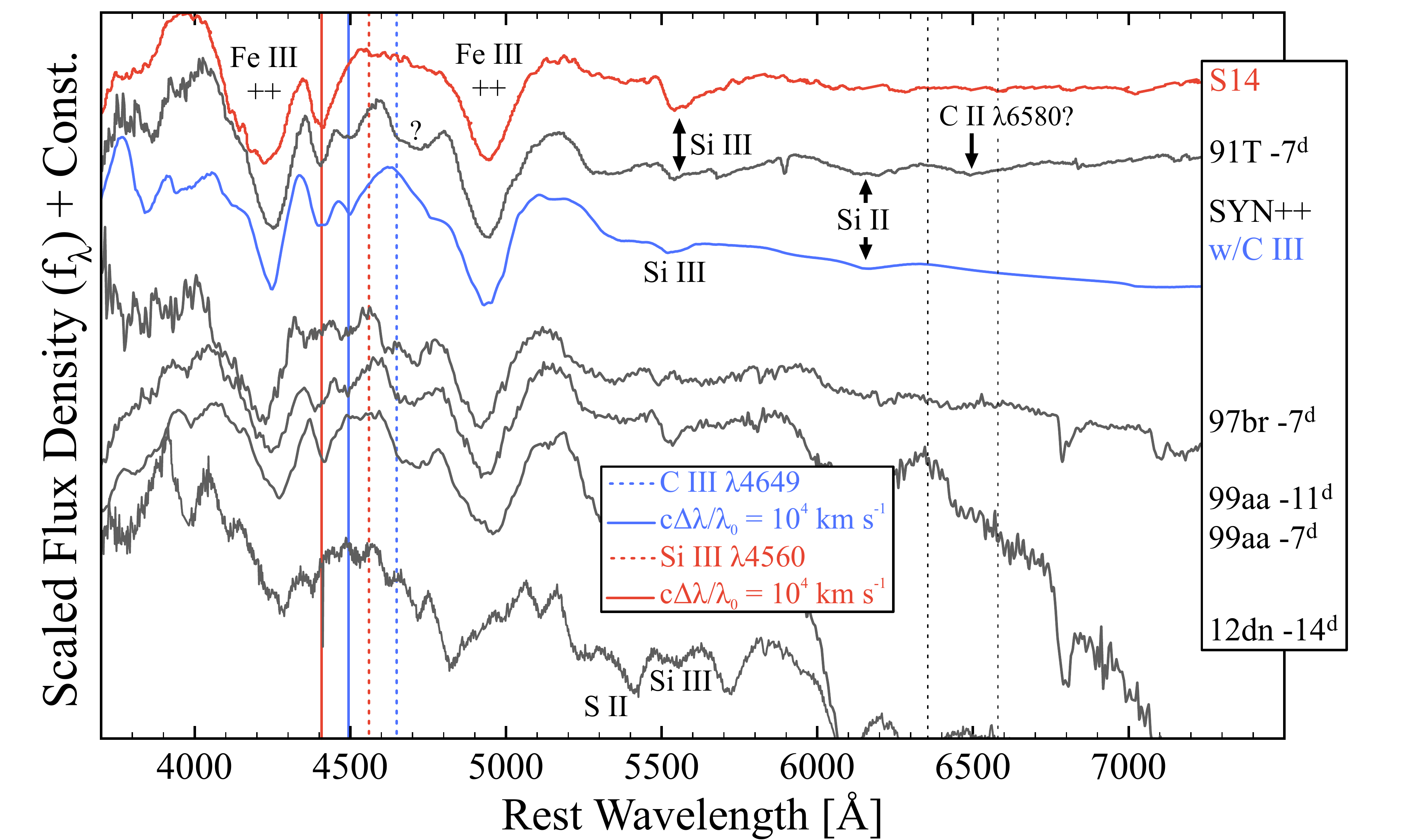} 
\caption{Comparisons between the delayed detonation spectrum from \citet{Sasdelli14} (S14; day $-$7); our preliminary \texttt{SYN++} spectrum with \ion{C}{III} included in the full fit (see text of \S6.4); and so-called ``Shallow Silicon'', SN~1991T/1999aa-like SN~Ia and SN~2012dn. Spectrum references: SN~1991T \citep{Mazzali95}; SN~1997br \citep{WLi99}; SN~1999aa \citep{Garavini04}. The dotted, vertical lines in black are centered at 6355~\AA\ and 6580~\AA, and the ``\ion{Fe}{III}++'' signifies ``more than \ion{Fe}{III}''.}
\label{Fig:sasdelli}
\end{figure*}

In the previous section we found \ion{C}{II} is not particularly favored as the primary contributor to the 4110 and 4550~\AA\ features. 
However, the degeneracy of candidate ions for these weak features is many-fold, and simply in terms of unburned carbon. 

For example, \ion{C}{I}~$\lambda\lambda$4772,~4932 may contribute to regions near 4550~\AA. Considering the low ionization potential of neutral carbon, one might expect minimal contribution from \ion{C}{I}~$\lambda\lambda$4772,~4932. Therefore simultaneous optical$+$NIR data and detailed modeling would be needed to investigate further \citep{Hoflich02,Marion15,Hsiao15}.

In addition, the region between 4450~\AA\ and 4550~\AA\ in the day~$-$14.6 spectrum of SN~2012dn could be influenced by \ion{C}{III}~$\lambda$4649 with a line velocity of~$\sim$~9600$-$13,000~km~s$^{-1}$. Given a relatively weak 4550~\AA\ signature, we did not include \ion{C}{III} in the prescription of ions used for SN~2012dn in the previous sections. However, \texttt{SYNAPPS} would favor \ion{C}{III} closer to \texttt{v$_{min}$}~=~12,000~km~s$^{-1}$ ($\pm$~1000~km~s$^{-1}$) given the consistency with values inferred from other species. More over, both \ion{C}{III} and \ion{S}{III} have similar ionization potentials, and this pairing has been suggested for SN~Ia such as SN~1991T, 1997br, and 1999aa \citep{Hatano02,Garavini04,Parrent11}. 

In Figure~\ref{Fig:sasdelli} we compare pre-maximum spectra of SN~1991T, 1997br, 1999aa, and 2012dn. We have also plotted the day~$-$7 spectrum from the delayed-detonation series computed by \citet{Sasdelli14} who concluded that the carbon-rich progenitor material in SN~1991T was burned out to $\sim$~12,500~km~s$^{-1}$, leaving no signatures of carbon (top of Figure~\ref{Fig:sasdelli} in red). In Figure~\ref{Fig:sasdelli} we also compare SN~1991T to a preliminary \texttt{SYN++} spectrum. With \texttt{v$_{min}$} set to 11,500~km~s$^{-1}$ ($\pm$~1000~km~s$^{-1}$) for each ion, the fit includes mostly \ion{Fe}{III}, but also minor contributions from \ion{C}{III}, \ion{Si}{II}, \ion{Si}{III}, \ion{S}{III}, \ion{Fe}{II}, \ion{Co}{III}, and \ion{Ni}{II} as these species do a fair job at reproducing most features. 

As was previously discussed by \citet{Hatano02}, \citet{Garavini04}, and \citet{Parrent11}, there is a distinct feature near 4500~\AA\ in the pre-maximum spectra of SN~1991T, 1997br, and 1999aa that could be significantly shaped, in part, by \ion{C}{III}~$\lambda$4649 forming near $\sim$11,500~km~s$^{-1}$. \citet{Sasdelli14} claim ``carbon lines are never detected'' for SN~1991T in spite of the well-placed weak signature near 4500~\AA, yet this feature is not reproduced in the computed spectra of \citet{Sasdelli14} for an alternative interpretation. 

This might imply that either the ionization balance for the model is inaccurate (possibly due to an assumed diffusive inner boundary; cf. \citealt{Blondin15}), or too much carbon has been burned in the model. Without more advanced analysis of SN~1991T spectra from \texttt{PHOENIX} and \texttt{CMFGEN}, we therefore cannot rule out an identification of \ion{C}{III} in these SN~1991T-like events.


\begin{figure}
\centering
\includegraphics*[scale=0.26, trim = 10mm 0mm 0mm 0mm]{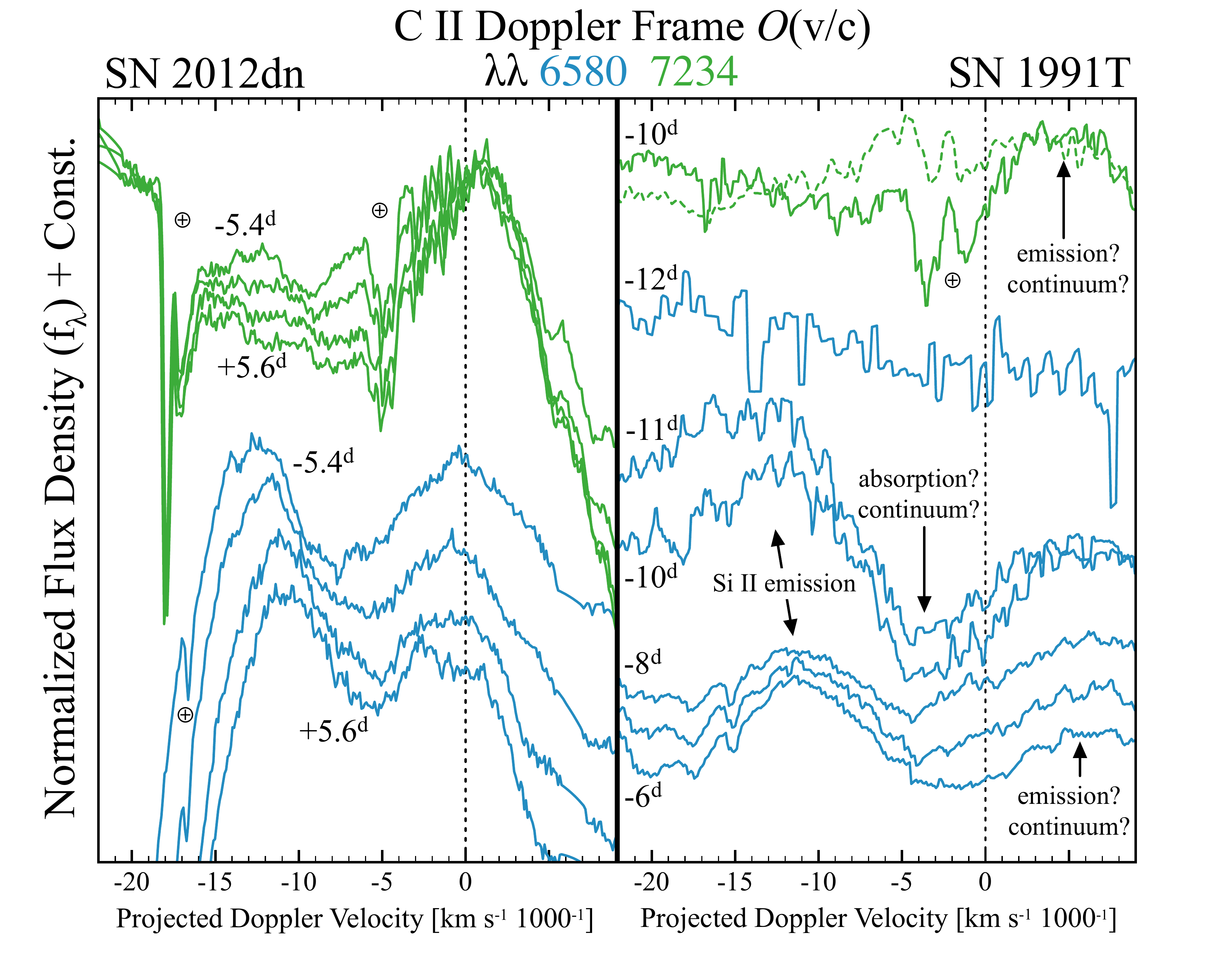} 
\caption{Time-series observations of SN~2012dn (this work) and 1991T \citep{Mazzali95} are compared in the Doppler frame $\mathcal{O}$(v/c) of \ion{C}{II}~$\lambda$6580 (blue) and $\lambda$7234 (green). In the right panel, the day $-$10 spectra from \citet{Mazzali95} (solid green) and \citet{Flipper91T} (dashed-green) are used to inspect the 7000$-$7200~\AA\ region of SN~1991T.}
\label{Fig:CIIframe}
\end{figure} 

Singly-ionized species might also account for some of the more weaker features observed for SN~1991T-like events. For example, by day~$-$7 for SN~1991T it is difficult to rule out weak contribution from \ion{Si}{II} near 6150~\AA. Notably, the computed spectrum of \citet{Sasdelli14} shown in our Figure~\ref{Fig:sasdelli} does not account for the 6150~\AA\ feature until three days later (see their Fig.~2). 

A more fleeting identification for SN~1991T$-$one that would further support the tentative detection of \ion{C}{III}~$\lambda$4649$-$is that of \ion{C}{II}~$\lambda$6580. Curiously, a weak depression red-ward of 6300~\AA\ is present in the day~$-$11 spectrum of SN~1991T and for sometime thereafter.

In Figure~\ref{Fig:CIIframe} we compare time-series spectra of SN~2012dn and 1991T in terms of line velocities $\mathcal{O}$(v/c) of \ion{C}{II}~$\lambda\lambda$6580, 7234 in order to access any common overlap between these two line signatures. This method was exploited by \citet{Parrent11}  for more normal SN~Ia, where it was found to be a good indicator of detectable \ion{C}{II} (see also \citealt{Thomas04}). Therefore, while we attempt to take advantage of common overlap in the Doppler frame in Figure~\ref{Fig:CIIframe}, we also emphasize that first-order estimates of projected Doppler velocities are not accurate, much less for solitary weak depressions in the spectrum that may simply be continuum (cf. \citealt{Branch77,Jeffery90}).

From the available data on SN~1991T, invoking a detection of weak \ion{C}{II}~$\lambda\lambda$6580, 7234 is troublesome. In particular, it is unclear where the would-be feature spanning 6350$-$6720~\AA\ ends and where the continuum begins. Furthermore, bonafide absorption signatures of \ion{C}{II}~$\lambda$6580 in normal SN~Ia are not often reported red-ward of $\sim$~6420~\AA\ \citep{Pereira13}.

By comparison, the bulk of the \ion{C}{II}~$\lambda$6580 signature in the day~$+$5.6 spectrum of SN~2012dn extends to $\sim$~6520~\AA, with a flat component extending toward $\sim$~6600~\AA. However, absorption from a parent line of \ion{C}{II}~$\lambda$6580 is not thought to be physical beyond a rest-wavelength of 6580~\AA\ during photospheric phases. Therefore our best interpretation is that the flat component centered about 6580~\AA\ is a result of either competing emission between \ion{C}{II} and \ion{Si}{II}, or an indication that some \ion{C}{II} is confined to a blob or shell-like structure (see \citealt{Jeffery90,Thomas07}).

As for SN~1991T, the feature nearest to 6580~\AA\ is clearly too far red-ward to be associated with \ion{C}{II} by day~$-$6. However prior to day~$-$7, what exactly is producing the minimum is unclear, yet the observations are not necessarily consistent with a non-detection of faint \ion{C}{II}. In Figure~\ref{Fig:CIIframe}, the day $-$10 spectra of SN~1991T from \citet{Mazzali95} (solid green) and \citet{Flipper91T} (dashed-green) are also too noisy to search for a weaker signature of \ion{C}{II}~$\lambda$7234. 


For SN~2012dn, where \ion{C}{II}~$\lambda$6580 is detected with a mean expansion velocity of $\sim$~12,500~km~s$^{-1}$, a weak 4450~\AA\ feature that can be reasonably pinned to \ion{C}{III}~$\lambda$4649 at $\sim$~12,500~km~s$^{-1}$ is only observed on day~$-$14.6, and is therefore unconfirmed as a feature. Thus, while we cannot rule out contribution from \ion{C}{III} near 4450~\AA, it would appear that signatures from \ion{C}{III} are minimal for SN~2012dn during photospheric phases.

\section{Discussion}

If a companion star is necessary for the runaway explosion of SN~Ia-like events (cf. \citealt{Chiosi15}), its identity and role in the diversity of post-explosion spectra remain fairly ambiguous \citep{Moll14,Maeda14,Tanikawa15}. Additionally, there are many scenarios that may explain spectroscopic diversity as projections of distinct progenitor systems. Extreme events, e.g. SN~1991bg and 2002cx, may result from mechanisms or progenitor channels that are distinct from those most representative of the true norm of SN~Ia (\citealt{Hillebrandt13} and references therein). A true norm could also originate from multiple binary configurations and explosive conditions \citep{Benetti05,Blondin12,Dessart14models}. Similarly, select subclasses may either represent bimodal characteristics of one or two dominant channels \citep{Maedanature,Maund10a,Roepke12,Liu15}, or a predominant family where continuous differences such as progenitor metallicities (among other parameters) yield the observed diversity \citep{Nugent95a,Lentz00,Hoflich02,Scalzo14a,Maeda14}. A dominant one (or two) progenitor system~$+$~explosion mechanism might then manifest properties about the norm that reach out into the same observational parameter space inhabited by the most peculiar subgroups \citep{Baron14}.



Nevertheless, an intriguing circumstance for SN~Ia is that the evolution of spectral features, and possibly the spectroscopic make-up, of Core Normal events can be considered similar, as a whole, to those of SN~2006gz, 2009dc, and 2012dn. Perhaps this would not be surprising since both classes of SN~Ia are expected to originate from WD progenitor systems. However, here we are referring to traceable similarities between Core Normal and SCC SN~Ia in the context of the contrasting spectroscopic make-up compared to peculiar SN~1991T-like, 1991bg-like, and 2002cx-like events \citep{Branch04a,Parrent11,Doull11}.\footnote{For comparison among proposed spectral sequences of SN, a far more diverse set of events, i.e. those of type IIbIbc and IIPLn, separately, are sometimes viewed as also forming a physically related sequence of progenitor configurations (\citealt{Nomoto96,Claeys11,Dessart12,Smith15}, see also \citealt{Maurer10}). For SN~IIbIbc, one of the key questions is how many ``hydrogen-poor SN~IIb'' are misidentified as ``hydrogen-poor SN~Ib'' when both events are primarily discovered and typed during a ``type Ib'' phase \citep{Chevalier10,Arcavi11,Danmil1311ei,Folatelli14,Parrent15}.}

\subsection{Does a unique physical sequence exist between normal and SN~2012dn-like events?}

If we assume that most normal SN~Ia yield similar rise-times ($\sim$~19~days), near-maximum light expansion velocities ($\sim$~11,000~km~s$^{-1}$), and opacities ($\kappa$~=~0.1~cm$^{2}$~g$^{-1}$), the values assumed for SN~2012dn (v$_{max}$~$\sim$~11,000~km~s$^{-1}$, t$_{rise}$~$\sim$~20 days; \citealt{Chakradhari14}) indicate M$_{ej}^{12dn}$~$\sim$~(90~$-$~120\%)~$\times$~M$_{Ch}$~=~1.3~$-$~1.7~M$_{\odot}$. ``By mass,'' SN~2012dn is not clearly SN~2009dc-like despite appearances (cf. \citealt{Pinto00b,Pinto00a,Foley09}).\footnote{If v$_{max}^{12dn}$~=~11,000~$\pm$~1500~km~s$^{-1}$, v$_{max}^{SNIa}$~=~12,000~$\pm$~2500~km~s$^{-1}$, t$_{rise}^{12dn}$~=~20~$\pm$~1.5 days, and t$_{rise}^{SNIa}$~=~19.3~$\pm$~0.5 days \citep{Conley06}, then M$_{ej}^{12dn}$~$\sim$~(70~$-$~130\%)~$\times$~M$_{Ch}$~=~1.0~$-$~1.8~M$_{\odot}$; the uncertainty is largely due to the poorly determined span of intrinsic mean expansion velocities. An additional uncertainty might arise from any difference between the mean opacities for normal and SN~2012dn-like events. See also \citet{Wheeler15}.} 
Furthermore, \citet{Chakradhari14} used Arnett's rule \citep{Arnett82} and log$L_{bol}^{peak}$~=~43.27~erg~s$^{-1}$ to estimate that SN~2012dn produced 0.70~$-$~0.94~M$_{\odot}$ of $^{56}$Ni. Assuming SN~2011fe and SN~2012dn have similar rise-times, \citet{Brown14} estimated the mass of $^{56}$Ni could be as low as $\sim$~0.5~M$_{\odot}$, where part of the uncertainty is in the rise-time of SN~2012dn.\footnote{\citet{Chakradhari14} estimate a lower-limit of 15.87 days.} By comparison, the bolometric luminosity of SN~2011fe is consistent with the production of $\sim$~0.42~$-$~0.64 M$_{\odot}$ of $^{56}$Ni \citep{Pereira13}, which is anywhere between half as much to slightly less than the amount of $^{56}$Ni inferred for SN~2012dn. 

For SN~2012dn, a significant fraction of the luminosity may be due to close circumbinary interaction with C$+$O-rich material. Therefore, the inferred mass of $^{56}$Ni needed to reproduce the observed rise-time would be artificially higher \citep{Taubenberger13}. \citet{Brown14} also report that the integrated luminosities of SN~2011aa, 2012dn (two SCC SN~Ia), and the normal and already NUV-blue SN~2011fe are quite similar despite the brighter UV luminosities for the SCC SN~Ia \citep{Milne12,Milne13}. Thus it seems plausible that SN~2011fe and SN~2012dn could have produced comparable amounts of $^{56}$Ni ($\pm$~0.10~$-$~0.20~M$_{\odot}$), where SN~2012dn obtained additional luminosity from close interaction ($\lesssim$~10$^{9}$~cm) with a larger exterior region of carbon-rich material. 

\begin{figure}
\centering
\includegraphics*[scale=0.28, trim= 130mm 0mm 40mm 0mm]{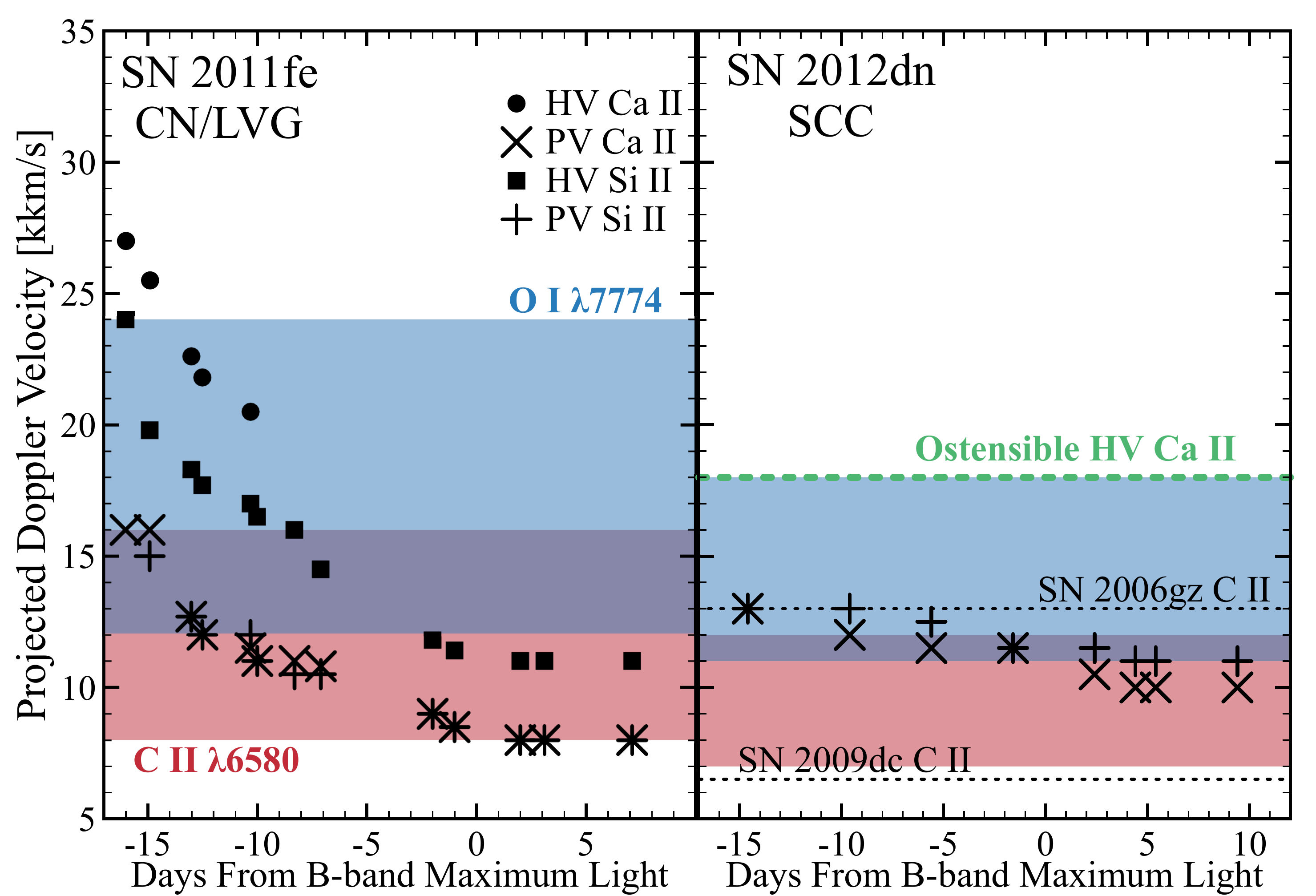}
\caption{Initial \texttt{SYNAPPS} measurements and comparisons between the ``Core Normal'', low-velocity gradient (CN/LVG) SN~2011fe and the SCC SN 2012dn. Symbols are defined in the left panel. Red and blue colors denote the span of inferred line velocities for \ion{C}{II}~$\lambda$6580 and \ion{O}{I}~$\lambda$7774, respectively, and the purple shade indicates overlap between the two. Owing to spectroscopic blending of weak identifying signatures, estimates for \ion{C}{II} and \ion{O}{I} are given as broad ranges corresponding to \ion{C}{II}~$\lambda$6580 and \ion{O}{I}~$\lambda$7774, respectively, rather than more well-determined v$_{min}$(t).}
\label{Fig:paradisefig}
\end{figure}

Our estimates of \texttt{v$_{min}$} for \ion{C}{II}, \ion{O}{I}, \ion{Si}{II}, and \ion{Ca}{II} for SN~2011fe and 2012dn are shown in Figure~\ref{Fig:paradisefig}. Notably, SN~2006gz, 2009dc, 2011fe, and 2012dn show a similar overlap between O, Si, and Ca-rich material, while carbon-rich regions for SCC SN~Ia share similar velocities found for Core Normal SN~Ia as well ($\sim$~11,000~$-$~15,000~km~s$^{-1}$). Compared to SN~2011fe, SN~2012dn only displays lower parameterized expansion velocities of PV \ion{Si}{II} during the initial stages of homologous expansion; i.e., PV~\ion{Si}{II} in SN~2011fe falls below that of SN~2012dn shortly thereafter between day~$-$10 and day~$-$5. 

For SN~2012dn, a non-detection of strong HV~\ion{Ca}{II} is obvious upon visual inspection of the 8200~\AA\ feature. However, we cannot rule out a faint signature of trace HV~\ion{Ca}{II} at $\sim$~19,000~km~s$^{-1}$ during pre-maximum epochs (\S6.2). If a region of detached HV \ion{Ca}{II} exists along the light-of-sight for SN~2012dn, then it is coincident with the lower extent of parameterized HV~\ion{Ca}{II} found in SN~2011fe (see Figure~\ref{Fig:paradisefig}).

Such considerations neither prove nor suggest that progenitor channels producing Core Normal and SCC SN~Ia subtypes are remotely identical, and similarly among other SN~Ia prototypes. Rather, they underscore the need for additional detailed investigations of spectroscopic degeneracies from single and double-degenerate scenarios \citep{Roepke12,Miles15,Dan15}.



\subsection{Differentiation Through Environmental Properties}

It may be possible to isolate candidate progenitor scenarios by noting the properties of the host-environment and host galaxy of a given event or SN~Ia subclass \citep{Hamuy95,Hicken09,WangX13,Pan14,Foley15,Maguire15}. The general finding has been that brighter, ``higher-stretch SN~Ia'' tend to occur in metal-poorer environments of low-mass host galaxies \citep{Sullivan10,Childress11,Khan11metalpoor}, while sub-luminous, ``low-stretch SN~Ia'' are preferentially detected in environments indicative of an older (or delayed) population of progenitor systems \citep{Howell01b,Gallagher08}.

However, this assumes that each spectroscopic subtype identically follow the above mantra, per host-property, which may not be the case \citep{Hicken09}. 
Specifically, the light curve stretch of one particular progenitor class or SN~Ia spectroscopic subtype, may tend to increase with metal-poor environments while another may not \citep{Kobayashi09,Kistler13}. In turn, this could serve as a means for separating progenitor scenarios or a sequence of progenitor masses for one dominant progenitor family \citep{Scalzo14a,Scalzo14subchandra}. However, the extent of such an effect remains unclear given the difficulty of classifying events from sporadically sampled spectra.

Below we use empirical relations found in the literature to place crude estimates on the host-galaxy mass and local metallicity to compare SN~2012dn to so-called nearest neighbor SN~Ia \citep{Jeffery07}. We also inspect the percentages of SN~Ia subclasses in the Palomar Transient Factory (PTF, \citealt{Rau09,Maguire14}) sample that was recently utilized by \citet{Pan14,Pan15} to investigate possible correlations between SN~Ia light curve stretch and local properties of the host environment. 

\subsubsection{Estimation of Host Galaxy Mass and Metallicity}

SN~2012dn is located 35'' west and 3'' south of the purportedly normal spiral (SAcd) galaxy, ESO~462-16 (=~PGC~64605; \citealt{Bock12}), which belongs to a pair of binary galaxies \citep{Soares95}. To estimate the host galaxy mass of SN~2012dn and hosts of other SN~Ia (\S6.2.2), we implement the empirical relation of \citet{Taylor11}:
\begin{equation}
logM/M_\odot = 0.70(g-i) - 0.40M_{I, AB} - 0.68
\end{equation}
and values obtained from NED and LEDA.\footnote{NED$-$http://ned.ipac.caltech.edu/; LEDA$-$http://leda.univ-lyon1.fr/. For M101, we assume log~$f_{\nu}$ (Jy) $\sim$ 0.45 at $\lambda$~$\sim$~0.806~$\mu$m.} 

Since our estimates here are not for precision cosmology, we assume representative errors by computing logM$_{Stellar}$~$\equiv$~logM/M$_\odot$ from Eq.~1 for the host of SN~2011fe, M101, and comparing to logM$_{Stellar}$~=~10.72$^{+0.13}_{-0.12}$ obtained by \citet{vanDokkum14}. Assuming {\it g}$-${\it i}~=~1.0, m$_{I, AB}$~$\sim$~7.8$\pm$0.2, DM~=~29.1$\pm$0.2 ({\it H}$_{0}$~=~69.7~km~s$^{-1}$ Mpc$^{-1}$; \citealt{Hinshaw13}) for M101, Eq.~1 yields logM$_{Stellar}$~$\sim$~8.6. Following the same procedure for the host of SN~2012dn, and taking the above discrepancy for M101 as a conservative error estimate, we obtain a broad window ($\pm$2.2) for all points in Figure~\ref{Fig:logM} below; i.e. approximately two orders of magnitude too broad to draw association with a particular progenitor environment or scenario for the present study.

We estimate the metallicity of SN~2012dn's progenitor from optical spectroscopy of the host environment. The SN flux dominates the host galaxy emission lines at the precise location of the explosion site, so we instead extract spectra from three nearby \ion{H}{II} regions which intersected our spectroscopic slit.  To model the emission line spectra and measure the fluxes of the strong emission lines, we apply the Markov Chain Monte Carlo method of \citet{Sanders12Ibc}. We then estimate metallicity using the N2 and O3N2 diagnostics of (\citealt{Pettini04}; hereafter, PP04).  Our spectra do not cover the [\ion{O}{II}]~$\lambda3727$ line, and we are therefore unable to estimate metallicity using diagnostics based on the $R_{23}$ line ratio.\footnote{$R_{23}$ $\equiv$ ([\ion{O}{III}~$\lambda$4959$+$5007] $+$ [\ion{O}{II}~$\lambda$3727])/H$\beta$ \citep{McGaugh91,Kewley02}.}

\begin{table}
 \centering
 \begin{minipage}{75mm}
  \caption{Metallicity Estimates of Literature SN~Ia}
  \begin{tabular}{ccc}
  \hline
SN & Approximate & Reference \\
Name & Metallicity (Z/Z$_{\odot}$) & Number \\
\hline
2003fg & 0.5 & \citet{Khan11metalpoor} \\
2006gz & 0.25 & \citet{Khan11metalpoor} \\
2007if & 0.15~$-$~0.20 & \citet{Khan11metalpoor} \\
& & \citet{Childress11} \\
2009dc & 0.38~$-$~0.52 & \citet{Silverman11} \\
2011fe & 0.3~$-$~0.5 & \citet{Mazzali14} \\
2012dn & 0.4~$-$~0.8 & this work \\
\hline
\end{tabular}
\end{minipage}
\end{table}

We find metallicity values largely consistent between the three studied regions, ranging between log(O/H)$_{\odot}~+~12~=~8.4 - 8.6$, or $\sim0.4 - 0.8~\rm{Z}_\odot$ on the PP04 scale ($\rm{Z}_\odot$~$\equiv$~log(O/H)$_{\odot}$~+~12~=~8.69; \citealt{Asplund05}).  The uncertainty in this estimate is small, $\sim0.05$~dex as estimated from propagation of the emission line flux uncertainties.  In particular, we find log(O/H)$_{\odot}~+~12~=~8.48~\pm~0.06$ in the PP04N2 diagnostic for the region nearest to the explosion site.  This uncertainty is likely to be dominated by systematic effects such as the calibration variance in the strong line metallicity diagnostics, local metallicity variation in the host galaxy, and chemical evolution and host environment physical offset associated with the delay time between the birth and explosion of the progenitor star (\citealt{Kewley08,Sanders12M31,Sanders12Ibc}, see also \citealt{Pan14,Toonen14,Soker14}).

On average, the metallicity inferred for the local environment of SN~2012dn places it at higher values relative to other SCC SN~Ia (see Table~3). If these values were accurate enough for comparisons (cf. \citealt{Niino15}), then at face value they would imply SN~2012dn originated from a progenitor with a metallicity no less than those giving rise to SN~2003fg, 2009dc, and the Core Normal SN~2011fe. However, it is only the metallicity estimate for SN~2011fe that stems from comparisons of UV spectra and abundance tomography (\citealt{Mazzali15,Baron15}). Thus the context by which to properly estimate and interpret progenitor metallicities with certainty remains absent \citep{Brown15}.

\subsubsection{Bulk Trends in SN~Ia Samples} 

Visual classification of SN spectra is sometimes described as a subjective or ambiguous process, whereby a computer is needed to facilitate the problem \citep{SasdelliMachine}.\footnote{This problem is not unique to supernova spectra. Solar spectroscopists, among others, have faced a similar debate, and we refer the reader to p.~23 of ``Stellar Spectral Classification'' \citep{Gray09} and  \citet{Keenan84} where visual classification of stellar spectra is discussed.} Between normal and peculiar events, or events within one subclass, this is somewhat true for noisy, post-maximum phase spectra, i.e., when spectroscopic diversity declines between most SN~Ia subtypes (\citealt{Branch09} and references therein). However, the unambiguous patterns established by pre-maximum light data indicate sub-classification can be done by-eye when the event is caught early enough and followed-up by campaigns of high quality observations. This is because noticeable differences of both major and minor features are often related to respective differences in the composition of the outermost regions of ejecta \citep{Tanaka08}.

With today's large spectrum-limited datasets, a computer is indeed ideal for charting dissimilarities between events in a given subclass, such as homogeneous Core Normal events most similar to SN~2011fe. This assumes, however, that some of the under-observed events in either a given sample or pre-defined subclass are not caught-late or ``peculiar'' outliers for a neighboring, and possibly unrelated, subclass such as Broad Line SN Ia (see also \citealt{Folatelli14,Parrent15}). In fact, there is curious overlap between events that have been typed as ``high velocity Core Normal'' and ``normal velocity Broad Line'' SN~Ia \citep{Blondin12,Parrent14}. Therefore, the vetting process for these and other subclasses of type I supernovae would benefit from spectral decomposition through gaussian processes \citep{Fakhouri15} and machine learning methods \citep{SasdelliMachine}.

\begin{figure}
\centering
\includegraphics*[scale=0.31, trim = 10mm 0mm 50mm 0mm]{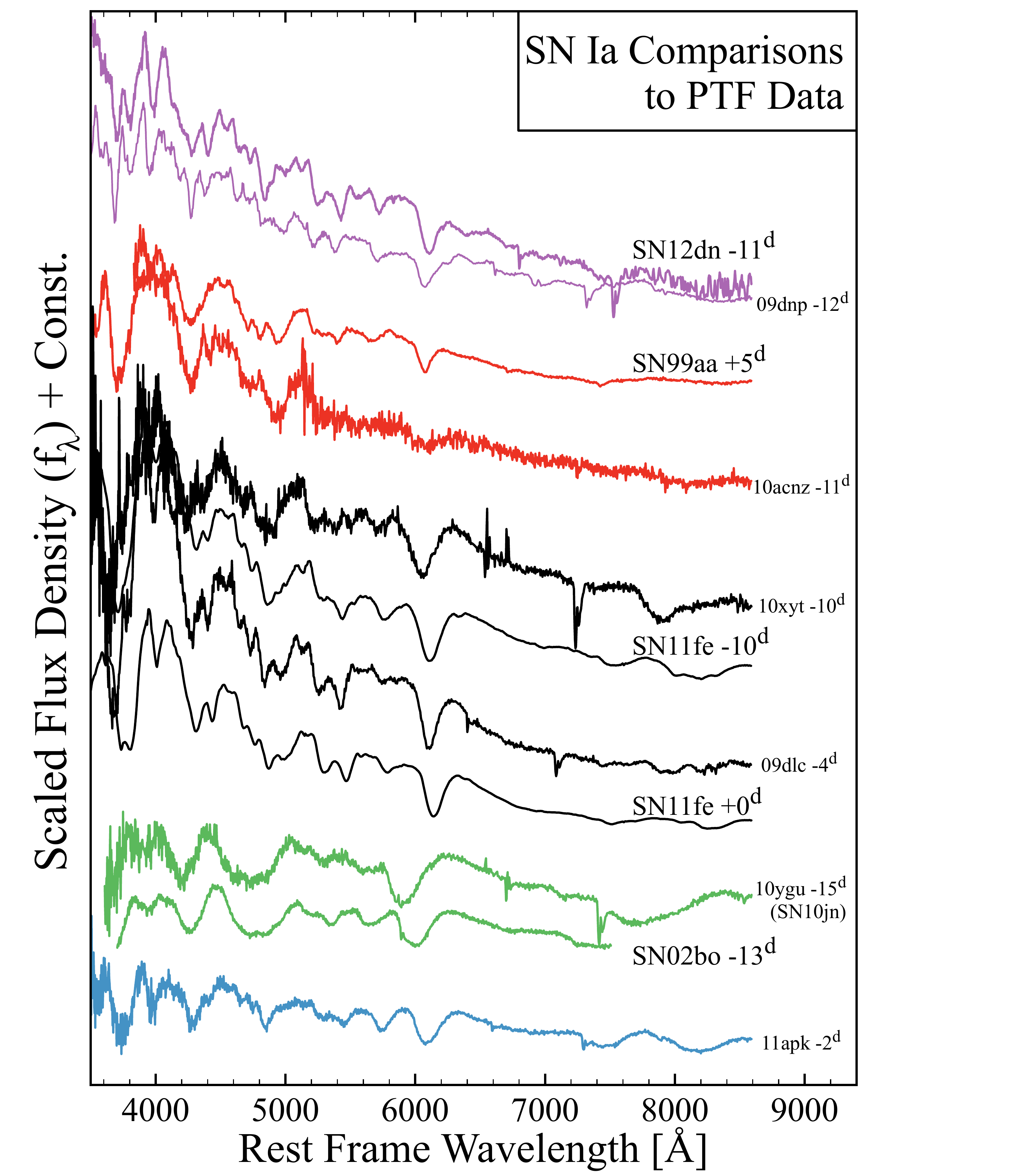} 
\caption{Comparisons between literature SN~Ia, SN~2012dn, and some of the best-observed PTF SN~Ia discussed in \citet{Pan14,Pan15}. The colors per subtype are in reference to the classification criteria of \citet{Branch06}; filled black circles for Core Normal, green for Broad Line, red for Shallow Silicon, blue for Cool, and purple for super-M$_{Ch}$ mass candidate events.}
\label{Fig:support}
\end{figure}

\begin{figure}
\centering
\includegraphics*[scale=0.46, trim = 530 80 150 40 ]{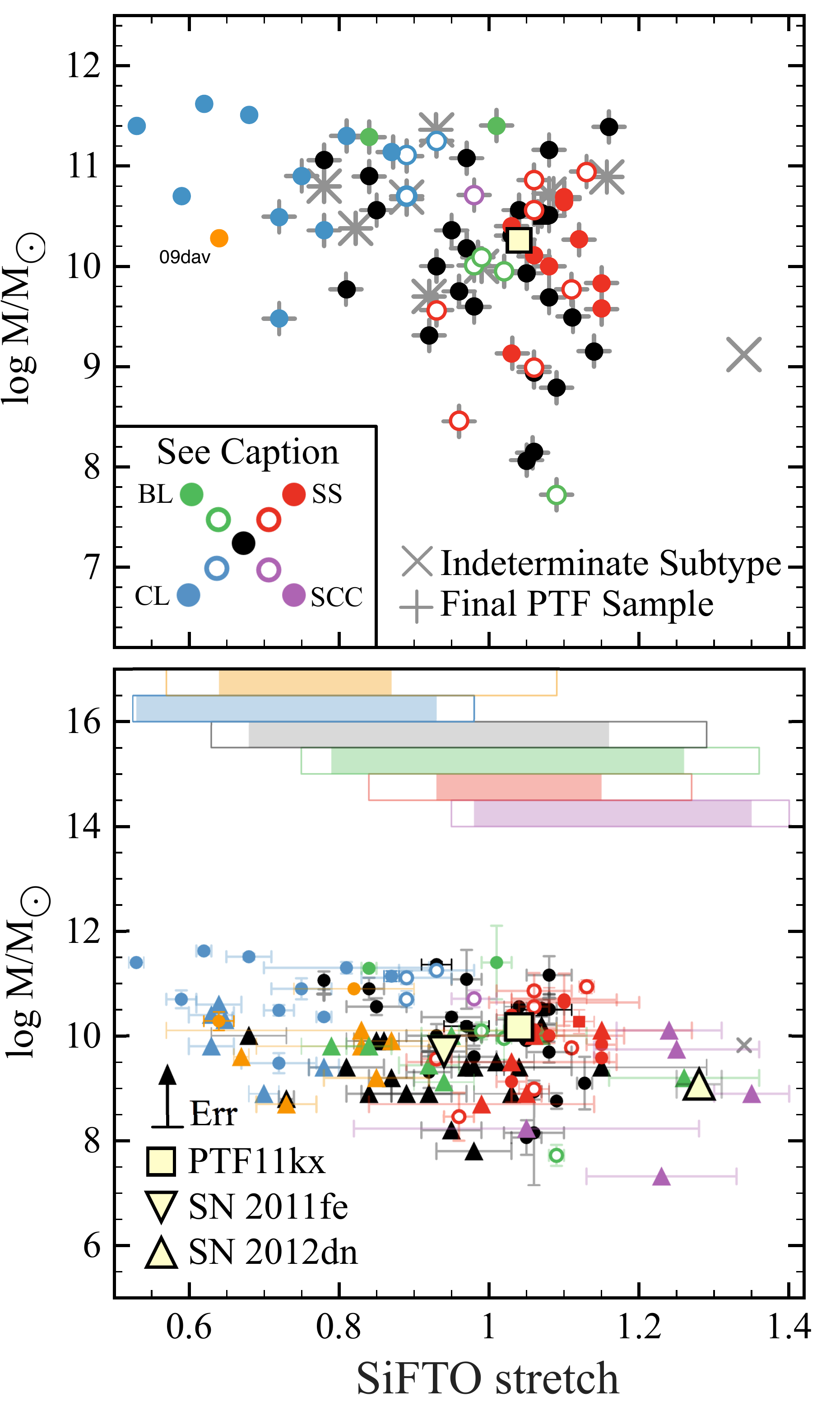}
\caption{[{\it Top}] host galaxy mass versus light curve stretch using the PTF sample SN~Ia initially presented by \citet{Pan14}, shown as grey $+$ symbols. Over-plotted in color are our best comparative guesses for the subtypes of each PTF SN shown. Colors are the same as in the adjacent legend, while orange circles denote Ia-peculiar events like SN~2002cx and PTF09dav. Open-circles indicate shared tendencies with a particular subtype. Grey $\times$s represent events with follow-up spectra too noisy or too late in epoch to visibly decipher a subtype. [{\it Bottom}] The above inset figure supplemented with literature SN~Ia and estimates for host galaxy mass using the empirical relation of \citet{Taylor11}. Horizontal bars represent the span of light curve stretch for each spectroscopic subtype shown. Here we are primarily concerned with the light curve stretch axis. However, for the purposes of plotting all literature SN~Ia, the triangles represent lower limit estimates of host galaxy mass as derived from Equation 1 and normalized to that for M101. Errors for PTF objects are retained from \citet{Pan14}. [{\it Note}] In a previous version of \citet{Parrent14}, the value for SN~2012dn's stretch in the plot did not match the correct value in \S3 of this work.}
\label{Fig:logM}
\end{figure}


From a sample of 82 PTF SN~Ia, \citet{Pan14} conducted a follow-up investigation of empirical relationships between SN~Ia light curve stretch and various properties of the host environment, including host galaxy mass, specific star formation rates, and gas-phase metallicities. Previous conclusions were reached insofar as the stellar mass of the host is correlated in the light curve shape of the subsequent explosion \citep{Sullivan10,Kim14}. However, as with most previous surveys of SN~Ia host-environments, visual details on the spectroscopic diversity of this most recent sample were omitted (see later \citealt{Pan15}).

In terms of signal-to-noise ratios and spectroscopic follow-up, it is important to note the data of the 82 PTF SN~Ia are of low to moderate quality. Similar to spectra obtained by the SuperNova Legacy Survey, PTF SN come with a few sporadically sampled optical spectra on average. As sub-classifications of supernovae are more accurately determined by time-series datasets, a lack of coverage inhibits a precise recording of those defining features most often associated with spectroscopic peculiarity, particularly during pre-maximum phases. For a visualization of this, \citet{REVIEW} presented a follow-up efficiency diagram in their Fig.~4. The PTF sample was included in this diagram, falling toward the bottom-left corner of their figure inset. 

Since we cannot attempt a consistent quantification of known subtypes for the PTF sample,  our assessments here, and similarly for the study of high velocity features by \citet{Pan15}, are based on visual comparisons to time-series observations of SN~Ia that have been designated as ``well-observed'' throughout the literature. Several examples of our by-eye sub-classifications are detailed in Figure~\ref{Fig:support}. 

To illustrate the distribution of SN~Ia subtypes for a given SN sample, we focus on the light curve stretch versus logM$_{Stellar}$ plane (see Fig.~9 of \citealt{Pan14}). In Figure~\ref{Fig:logM} we have replotted the PTF sample as grey $+$ symbols. The colors per subtype are in reference to the classification criteria of \citet{Branch06}; filled black circles for Core Normal, green for Broad Line, red for Shallow Silicon, blue for Cool, and purple for super-Chandrasekhar mass candidate events. For PTF SN~Ia that appear normal but are suspected as having properties of neighboring subtypes, e.g., Core Normal, Broad Line, and a selection of peculiar hybrids such as SN~1999aa, we have marked these as colored open-circles where appropriate, and we consider this as an error estimate below. Objects for which a spectroscopic subtype could not be determined from an appropriate time-series dataset are marked as grey $\times$s. A number of these cases are related to an inability to disentangle normal velocity SN~1984A-like from high[er] velocity SN~1994D-like events, even after enforcing an artificial selection criterion such as a maximum light separation velocity of 11,800~km~s$^{-1}$ \citep{WangX09Subtype} or 12,200~km~s$^{-1}$ \citep{Blondin12}.

In Table~4 we list estimates of SN~Ia subtype percentages for the sample in terms of \texttt{SiFTO} light curve stretch and find considerable room for additional trends, per subtype. Here, the overall trend between logM$_{Stellar}$ and light curve stretch is built from a heterogenous aggregate of visibly distinct SN~Ia, with traceable peculiarities, that may or may not be physically related to other spectroscopic subtypes included in the sample. As such, the claim of ``higher stretch SN~Ia originate from low mass, metal-poor environments'' is not explicitly with regards to Core Normal, SN~2011fe-like SN~Ia with light curve stretch values between 0.8 and 1.3 (see Figure~\ref{Fig:logM}). \citet{Pan14} do not claim otherwise, however we find it important to explicitly show these details when addressing bulk trends pertaining to, e.g., properties of the host.


The percentages of each subtype in the PTF sample are also distinct from those found in the volume-limited sample of \citet{WLi11b}, where SN~Ia were classified by likeness to a few historical prototypes (cf. \citealt{Flipper97} and Fig.~13 of \citealt{Doull11}). We suspect a quantitative re-evaluation of SN~Ia subtypes would better elucidate the multitude of trends involving SN~Ia rates and host-galaxy properties. 

\subsubsection{Correlations with \ion{Ca}{II} at High Velocities}

\citet{Pan15} later analysed properties of \ion{Si}{II}~$\lambda$6355 and the \ion{Ca}{II} near-infrared triplet profiles for 122 SN~Ia and found trends relating to the stellar mass and star formation rates of the host galaxy. In particular, \citet{Pan15} recover a trend, previously found by \citet{Maguire12,Maguire14} and \citet{Childress14}, where maximum light spectra that have strong signatures of HV~\ion{Ca}{II} tend to also have broader light curves. All of these works deduce the relative strengths and velocities (weighted by the blended absorption minimum) of high velocity features near maximum light using up to five gaussian absorption profiles to encompass 11 P~Cygni-like lines; i.e., effectively narrow-band pass-filters with one gaussian for PV \ion{Si}{II}~$\lambda$6355 and four gaussians for PV and HV~\ion{Ca}{II} to represent two unconstrained components of H\&K and the infrared-triplet.

During pre-maximum epochs, such methods can break down without the inclusion of either parameterized HV~\ion{Si}{II}, or an extended region of \ion{Si}{II} in absorption (cf. \citealt{Blondin15}). Furthermore, it is not often that signatures of detached HV~\ion{Si}{II} are as prominently observed as in the case of the peculiar SN~2012fr \citep{Childress13}, nor does an interpretation of distinctly detached shells of HV~\ion{Si}{II} and HV~\ion{Ca}{II} necessarily follow quantitative consistency between observations and a prescription of PV~$+$~HV~\ion{Si}{II} and \ion{Ca}{II} \citep{Baron15}.

Additionally, parameterized HV components of \ion{Si}{II} and \ion{Ca}{II} inferred for the normal SN~Ia~2011fe are not coincident in such a way that one can readily deduce the outermost extent of \ion{Ca}{II} from that of \ion{Si}{II} (see our Figure~\ref{Fig:paradisefig}). Moreover, the estimated expansion velocities for a detached component of HV~\ion{Ca}{II} can be uncertain by 2000$-$8000~km~s$^{-1}$ depending on the width of the feature and the observational coverage (see Fig~3 of \citealt{Parrent14}). There has also been evidence for multiple shells of high velocity \ion{Ca}{II}, as suggested for SN~2000cx (\citealt{Thomas04}, see also \citealt{Silverman13SN2013bh}).

Weak signatures of \ion{C}{I}, \ion{O}{I}, \ion{Mg}{II}, and \ion{Fe}{II} might also influence wavelengths that undergo absorption from the \ion{Ca}{II} infrared-triplet. For events where there are sufficiently early pre-maximum spectra, e.g., SN 1994D and 2011fe, signatures of HV~\ion{O}{I} may be present as well \citep{Branch05,Nugent11,Parrent12}. Near maximum light, one might also expect increased absorption from \ion{O}{I} for fast-decliners such as SN~1991bg, 1997cn, 1999by, 2000dk, 2005bl, and 2007fr (\citealt{Doull11} and references therein). More significant contributions may come from lines of \ion{Co}{II} \citep{Blondin15}. This translates to possibly overestimating contributions from parameterized PV~\ion{Ca}{II} and underestimating the extent of trace \ion{Ca}{II} at relatively higher velocities, which can go unconfirmed without observations prior to $\sim$~day~$-$7.


\citet{Silverman15} recently improved upon the methods of \citet{Maguire12,Maguire14}, \citet{Childress14}, and \citet{Pan15} by considering {\it gf}-weighted line-strengths. \citet{Silverman15} also test their method by fitting 11 out of 445 spectra with \texttt{SYNAPPS}. Some of their preliminary fits include PV components of \ion{O}{I}, \ion{Mg}{II}, \ion{Si}{II}, \ion{Si}{III}, \ion{S}{II}, \ion{Ti}{II}, \ion{Fe}{II}, \ion{Fe}{III}, and an additional HV component of \ion{Ca}{II}.

Using \texttt{SYNAPPS} in this manner can effectively confirm the need for parameterizing HV \ion{Si}{II} and HV~\ion{Ca}{II}, however it does not support an accurate determination of parameterized velocities from gaussian fitting methods where fewer ions are incorporated. This is primarily because the broad features of supernova spectra enable a gross approximation of semi-empirical quantities, which are dependent on optical-depth effects and ionization gradients in the outermost regions of ejecta. 

On this point, the six fits shown in Fig.~1 of \citet{Silverman15}, and particularly for SN~1994D, 2002dj, and 2012fr, indicate either an over/under-shooting of velocities, or overly saturated lines. \citet{Silverman15} do in fact attribute some of the discrepancies in their fits to the degeneracy between \texttt{SYNAPPS} parameters, \texttt{v$_{min, Y}$} and \texttt{logtau}. Although, \citet{Silverman15} do not show the simultaneous fit to the H\&K and infrared-triplet region for any event (see also Fig.~1 of \citealt{BenAmi15} and Fig.~17 of \citealt{Pignata11}), which is important for zeroing-in on self-consistent estimates of \texttt{v$_{min, Y}$}. Moreover, degeneracies between \texttt{SYNAPPS} parameters can be broken from time-series fitting, and can also result in an improved time-dependent (and model-dependent) accuracy of a supposed separation velocity.

Still, studies that have attempted a complete tracing of SN~Ia spectral features with \texttt{SYNAPPS} do not unanimously produce estimates of v$_{min}$ better than $\sim$~$\pm$1000~km~s$^{-1}$ (\citealt{Parrent12,Parrent14}; and this work). Overall, the direct analysis methods of \citet{Parrent12}, \citet{Maguire12,Maguire14}, \citet{Childress14}, \citet{Pan15}, and \citet{Silverman15} are so far unable to determine when a broad feature is due to a detached (and blended) component at higher velocities, and when the width of that feature is produced by a changing ionization gradient. These methods do give credence to the inclusion of parameterized HV components, but the inferred separation velocities remain uncertain outside of events similar to SN~2012fr.

Near maximum light, it is true that most events no longer harbor a strong feature near 8000~\AA, or at least not similar to those of SN~1999ee, 2001el, and 2004S (which do not appear to have been included in the studies done by \citealt{Childress14} and \citealt{Silverman15}). However, the inclusion of Broad Line events in the sample of \citet{Childress14}, such as SN~2006X, is debatable considering that line velocities for the \ion{Ca}{II} infrared-triplet in SN~2006X overlap those of HV~\ion{Ca}{II} in some normal SN~Ia (16,000$-$25,000~km~s$^{-1}$), yet some have interpreted SN~2006X as having no high velocity features because there are no distinguishable signatures of HV~\ion{Ca}{II} upwards of 20,000$-$30,000~km~s$^{-1}$ prior to maximum light (see also \citealt{Silverman15}).

For SN~Ia that exhibit spectral features of high ionization lines, e.g. SN~1991T, 1998es, and 1999aa, the bulk of their infrared \ion{Ca}{II} feature appears to peak in strength $\sim$~1-week after maximum light. This is still consistent with a lack of strong signatures of HV \ion{Ca}{II} near maximum light, yet \citet{Silverman15} find favorable evidence for weak absorption from \ion{Si}{II} and \ion{Ca}{II} at nominally high velocities for some of these Shallow-Silicon events.

From a sample of 18 SN~1991bg-like events with coverage near maximum light, \citet{Silverman15} conclude models of SN~1991bg should never produce high velocity signatures. However, this too remains uncertain given that the only SN~1991bg-like event with a spectrum taken as early as day~$-$6 is SN~2005bl \citep{Taubenberger08,Hachinger09}, and the normal SN~Ia~2011fe effectively lost conspicuous signatures of highly blue-shifted absorption in \ion{Ca}{II} near day~$-$10.

As for the positive correlation between the strength of HV~\ion{Ca}{II} near maximum light and the width of the light curve, there are SN~Ia with broad light curves that do not also have strong, high velocity features of \ion{Ca}{II}, e.g., SN~2006gz, 2007if, and 2009dc (\citealt{Silverman15} and references therein). Yet from our analysis of SN~2012dn, we are still unable to directly rule out faint signatures of detached \ion{Ca}{II} from its spectra using the relatively stronger H\&K lines superimposed with \ion{Si}{II}~$\lambda$3858. Whether such a detection is physically realizable underscores the need for refining parameter ranges for \texttt{SYNAPPS}-like tools by calibrating to spectrum synthesizers that incorporate an approximated abundance model, e.g., \texttt{TARDIS} \citep{Kerzendorf14}.

\section{Conclusions}

The time-series spectra we obtained of SN~2012dn during its first month showed an evolution most similar to SCC SN~Ia, but not as luminous as one might expect \citep{Brown14,Chakradhari14}. In particular, the weaker than normal signatures of \ion{Si}{II} and \ion{Ca}{II}, a stronger than normal signature of \ion{C}{II}~$\lambda$6580, and relatively slowly evolving line-velocities indicate that SN~2012dn is most closely associated with the SCC class of SN~Ia. Late-time comparisons to SCC SN~Ia also support an association between this particular subclass and SN~2012dn (see \citealt{Chakradhari14} and our Figure~\ref{Fig:comp}). 


The 11-component prescription of ions utilized here for deconstructing the photospheric phase spectra of SN~2012dn (Figures~\ref{Fig:comps_0730}$-$\ref{Fig:carboncomps}) afforded us an ability to provide internally consistent inferences of multiple line-velocities (Figure~\ref{Fig:paradisefig}), and can be improved upon \citep{Parrent14}. Compared to signatures of \ion{C}{II} observed for the normal SN~2011fe, for SN~2012dn we estimate the material associated with \ion{C}{II} is only $\sim$~2000~km~s$^{-1}$ slower.  As was similarly proposed for SN~2006bt \citep{Foley10}, \citet{Chakradhari14} find the discordant velocity estimates of \ion{C}{II} and \ion{Si}{II} may indicate that the carbon-rich material is clumpy and off-center from the line-of-sight. Whether or not this is the case awaits a larger sample of SN~2012dn-like events. Additionally, and unlike other so-called ``Shallow Silicon'' events such as SN~1991T, 1997br, and 1999aa, we find little evidence favoring a detection of \ion{C}{III}~$\lambda$4649 in the early phase spectra of SN~2012dn.

An important take away point from our analysis of SN~2012dn spectra in \S6 and visual classification of PTF SN~Ia in \S7 is that a majority of SN spectral features are not caused by some underlying stochastic variation. Rather, signatures of atomic species form a highly blended spectrum that nature is frequently capable of reproducing, such that we have a few distinct SN~Ia subtypes and a number of outliers that suggest a continuum. Yet a cross examination of spectra computed for super-M$_{Ch}$ merging WDs (e.g., \citealt{Pakmor12,Moll14,Raskin14}) reveals that, while these spectra resemble SN~Ia-like events, they are not well-matched to observations of super-M$_{Ch}$ candidates SN~2006gz and 2012dn, nor have they been shown to consistently predict the spectroscopic evolution of more normal SN~Ia.

Namely, \citet{Moll14} and \citet{Raskin14} find line velocities that are too high for both normal SN~Ia and events similar to SN~2012dn (upwards of 20,000~km~s$^{-1}$ near maximum light, and depending on the viewing-angle).  Similarly, the maximum-light spectrum for the violent merger provided by \citet{Pakmor12} has the appearance of predicting signatures of iron that are too strong compared to normal SN~Ia, whereas signatures of intermediate-mass elements do not appear strong enough. Given that signatures of iron-peak and intermediate-mass elements in SN~2006gz and 2012dn are somewhat weaker than normal, and therefore weaker than some of the latest predictions of merging WDs, we are therefore unable to relate these progenitor scenarios in their current form to SN~2012dn. 

In an attempt to differentiate a progenitor channel through other means, in \S7 we examined the PTF sample used by \citet{Pan14} who studied correlations between properties of the host galaxy and light curve parameters of 82 SN~Ia. However, in \S7.2.2 we found that empirical measures such as light curve width, or stretch, do not encompass the spectroscopic details for a given sample of SN~Ia (see Figures~\ref{Fig:support}~and~\ref{Fig:logM}), nor has the negative correlation between light curve stretch and host-galaxy mass been shown to apply to all types of peculiar and more normal SN~Ia  within a given sample. 

Current models of both single-degenerate delayed detonations \citep{KasenPlewa07} and select WD mergers \citep{Pakmor12,Raskin14,Dan15} may account for multiple SN~Ia subtypes as the distribution of ejected material is found to be asymmetric in some instances. Therefore, to investigate and clarify the relationships between spectroscopic diversity, properties of the host environment, and lifetimes of the progenitors, photometrically-based SN surveys would benefit from enhanced spectroscopic monitoring of nearby supernovae, including events at redshifts greater than unity.

 \begin{table*}
 \centering
 \begin{minipage}{175mm}
 \centering
  \caption{Estimates for SN~Ia Subtype Contamination in the ``Quality Cut Sample'' from \citet{Pan14}}
  \begin{tabular}{lrc}
  \hline
SN~Ia Subtype & Sample Percentage & \texttt{SiFTO} Stretch \\
\hline
Core Normal, SN~1994D, 2011fe, 2014J-likes & 38$-$58\% & 0.78~$<$~s~$<$~1.16 \\
Broad Line; SN~1984A-, 2002bo-likes & 3$-$9\% & 0.84~$<$~s~$<$~1.09 \\
Shallow Silicon, SN~1991T and intermediary 1999aa-likes & 14$-$24\% & 0.93~$<$~s~$<$~1.15 \\ 
Cool, SN~1991bg and intermediary 2004eo-likes & 9$-$13\% & 0.72~$<$~s~$<$~0.93 \\
Ia-CSM (PTF11kx) and super-Chandrasekhar candidate (PTF09dnp) & 3\% & s = 1.04 and 0.98, respectively \\
\hline
\end{tabular}
\end{minipage}
\end{table*}

\section*{Acknowledgements}

This work was supported by the Las Cumbres Observatory Global Telescope Network. This research used resources of the National Energy Research Scientific Computing Center, which is supported by the Office of Science of the U.S. Department of Energy under Contract No. DE-AC02-05CH11231, and a grant from the National Science Foundation: AST-1211196. JV is supported by Hungarian OTKA Grant NN 107637. 

We are indebted to Asa Bluck, Ruben Diaz, Peter Pessev, Kathy Roth, and Ricardo Schiavon for their critical role in obtaining the Gemini Target-of-Opportunity data with both Gemini facilities (PI, D.~A.~Howell, GN-2012A-Q-34 and GS-2012A-Q-20; processed using the Gemini IRAF package), which are operated by the 
Association of Universities for Research in Astronomy, Inc., under a cooperative agreement 
with the NSF on behalf of the Gemini partnership: the National Science Foundation 
(United States), the National Research Council (Canada), CONICYT (Chile), the Australian 
Research Council (Australia), Minist\'{e}rio da Ci\^{e}ncia, Tecnologia e Inova\c{c}\~{a}o 
(Brazil) and Ministerio de Ciencia, Tecnolog\'{i}a~e~Innovaci\'{o}n Productiva (Argentina). Observations reported here were also obtained at the MMT Observatory, a joint facility of the Smithsonian Institution and the University of Arizona. Additional observations reported in this paper were obtained with the Southern African Large Telescope.

This work was also made possible from contributions to the SuSupect \citep{Richardson01} and WISeREP databases \citep{WISEREP}, as well as David Bishop's Latest Supernovae page \citep{Galyam13}, and has made use of PTF data obtained and reduced by J.~Cooke, S.~Ben-Ami, and R.~S.~Ellis. JTP wishes to thank D.~Branch, E.~Baron, and A.~Soderberg for helpful discussions and comments on previous drafts. 

Finally, we wish to thank our anonymous referee for helpful comments on a earlier draft. 

{\it Facilities:} Las Cumbres Global Telescope Network Inc., Gemini, FTS, FLWO, SALT, MMT.

\bibliographystyle{mn2e}
\bibliography{jparrent_bib}{}




\label{lastpage}

\end{document}